\documentclass[reprint, aps, amsmath, showpacs, superscriptaddress]{revtex4-1}
\usepackage{graphicx}
\usepackage{epsfig}
\usepackage[cp1251]{inputenc}
\usepackage[english]{babel}
\usepackage{amsmath}
\usepackage{amssymb}
\usepackage{amstext}
\usepackage{color}
\usepackage{float}
\usepackage{epstopdf}
\usepackage[colorlinks=true,linkcolor=blue,allcolors=blue]{hyperref}
\usepackage{csquotes}

\begin{document}

\title{Modified rigorous coupled-wave analysis for grating-based plasmonic structures with delta-thin conductive channel. Far- and near-field study.}

\author{Yurii M. Lyaschuk}
\affiliation{Department of Theoretical Physics, Institute of Semiconductor Physics NAS of Ukraine, pr. Nauki 45, 03028 Kyiv, Ukraine}
\author{Serhii M. Kukhtaruk}
\affiliation{Department of Theoretical Physics, Institute of Semiconductor Physics NAS of Ukraine, pr. Nauki 45, 03028 Kyiv, Ukraine}
\affiliation{Experimentelle Physik 2, Technische Universit\"{a}t Dortmund, Otto-Hahn-Str. 4a, 44227 Dortmund, Germany}
\author{Vytautas Janonis}
\affiliation{Center for Physical Sciences and Technology, Saul\.{e}tekio al. 3, LT-10257 Vilnius, Lithuania}
\author{Vadym V. Korotyeyev}
\email{vadym.korotieiev@ftmc.lt}
\affiliation{Center for Physical Sciences and Technology, Saul\.{e}tekio al. 3, LT-10257 Vilnius, Lithuania}

\begin{abstract}
The modified rigorous coupled-wave analysis technique is developed to describe the optical characteristics of the plasmonic structures with the grating-gated delta-thin conductive channel in the far- and near-field zones of electromagnetic waves. The technique was applied for analysis of the resonant properties of AlGaN/GaN heterostructures combined  with deeply-subwavelength metallic grating
which facilitates the excitation of the two-dimensional plasmons in the THz frequency range. The convergence of the calculations at the frequencies near the plasmon resonances is discussed.
The impact of the grating's parameters, including filling factor and thickness of the grating, on resonant absorption of the structure was investigated in details.
The spatial distributions of electromagnetic field in a near-field zone were used for the evaluation of total absorption of the plasmonic structures separating contributions of the grating-gated two-dimensional electron gas and the grating coupler.
\end{abstract}

\maketitle

\section{Introduction}
Nowadays, structures with spatially-periodical lateral structurization/metasurfaces  are in  a focus of studies as the key elements of the many opto- and optoelectronics devices with broad application areas, including spectroscopy, imaging, holography, optical lithography, biochemical sensing and military application (for more information, see recent review in Ref.~\cite{Popov_Grating}). For example, diffractive gratings are widely utilized in different spectral ranges from microwaves to deep ultraviolet as dispersive optical component~\cite{Neumann}, antenna elements~\cite{Antenna}, polarizers~\cite{Polarizers}, etc.

Recently, great attention has been paid to exploitation of the subwavelength metasurfaced structures in THz and Far-infrared spectral ranges. Particularly,  the semiconductor structures with surface-relief gratings~\cite{Shaligin2016, Vitovt2020}, quantum well (QW) heterostructures~\cite{Popov,Korot2018} or graphene-based structures~\cite{Rizhii2020} incorporated with metallic gratings are widely discussed as potential efficient emitters and detectors of the THz radiation~\cite{Otsuji2014}. In such structures, the gratings play a role of the coupler, facilitating the resonant interaction between incident electromagnetic ($em$) waves and charge density waves such as surface plasmon-polaritons or 2D plasmons. Moreover, detailed investigations of the resonant properties of grating-based plasmonic structures can serve as additional tool for basic characterization of the 2D electron gas (2DEG) in QWs~\cite{Pashnev2020, Pashnev2020b}, graphene~\cite{Yan2015, Jadidi2015, Zhao2015, Lu2016, Kukhtaruk} and other novel 2D layered materials~\cite{Low}.

Mentioned research is faced with the problem of the rigorous electrodynamic simulations of optical characteristics of grating-based structures containing
very thin or even atomically thin conductive layers. In the simulations, these layers should be treated as delta-thin with two-dimensional parameters such as the sheet conductance of the electron channel. For such structures, the technique of integral equations (IE) is often applied for the solutions of the Maxwell's equations. This technique uses Green function formalism and is based on an reduction of the Maxwell's system of equations to the linear integral equations. Latter can be solved, for example, using Galerkin schemes with guaranteed convergence. The IE technique has been exploited
for the different problems including investigations of 2D plasmon instabilities under the grating~\cite{Michailov,Korot2018,Korot2020}, detection of THz radiation~\cite{Popov2010, Popov2011, Korot2017} and interaction of THz radiation with conductive-strip gratings~\cite{Nosich2013Aip, Nosich2013}.
In spite of apparent advantages with respect to fast convergence, this method is typically formulated for modeling structures with simple geometries and grating is treated as delta-thin.

Another powerful method for solution of grating-related electrodynamic problem is known as rigorous coupled-wave analysis (RCWA)~\cite{Moharam, Moharam1995, Gaylord}. This method belongs to the matrix-type methods and operates with systems of algebraic equations which are formulated for coefficients of the Fourier expansion of the actual components of the $em$ field. The RCWA can be applied for arbitrary complexity of the grating-based structures with any geometry of gratings. Typically, the realization of the RCWA relates to the structures where each layer is described by the bulk parameters. The RCWA method for structures with delta-thin conductive channels requires essential modifications which were recently discussed in Refs~\cite{Inampudi2017, Inampudi2019} for plasmonic structures with graphene. The aim of this paper is to present such modifications in more detail, focusing on study of convergence of the proposed method on example of calculations of optical characteristics of metallic grating-based resonant plasmonic structure with QW in THz frequency range. Effect of the grating depth on the plasmon resonance, analysis of the near-field pattern, and comparison to other methods are discussed. A total absorption of the plasmonic structures separating contributions of the grating-gated 2DEG and metallic grating couplers is also simulated using spatial distributions of the $em$ field in the near-field region.

Mathematical formalism of modified RCWA method is presented in Section \ref{Sec1}. RCWA will be formulated for planar diffraction problem (plane of incidence is perpendicular to grating strips) including the cases of TM and TE polarizations. The investigations of the convergence of proposed method and comparison with IE technique will be illustrated in Section \ref{Sec2} on example of the calculations of transmission, reflection and absorption spectra of QW plasmonic structure with deeply subwavelength metallic grating. The effect of the finite thickness of the metallic grating will be studied in details.
In Section \ref{Sec3}, we will perform the analysis of the near-field patterns under conditions of the plasmon resonances.
The main results will be summarized in Section \ref{Sec4}.

\section{Mathematical formalism}\label{Sec1}
\begin{figure}[htbp]
\centering
\fbox{\includegraphics[width=0.4\textwidth]{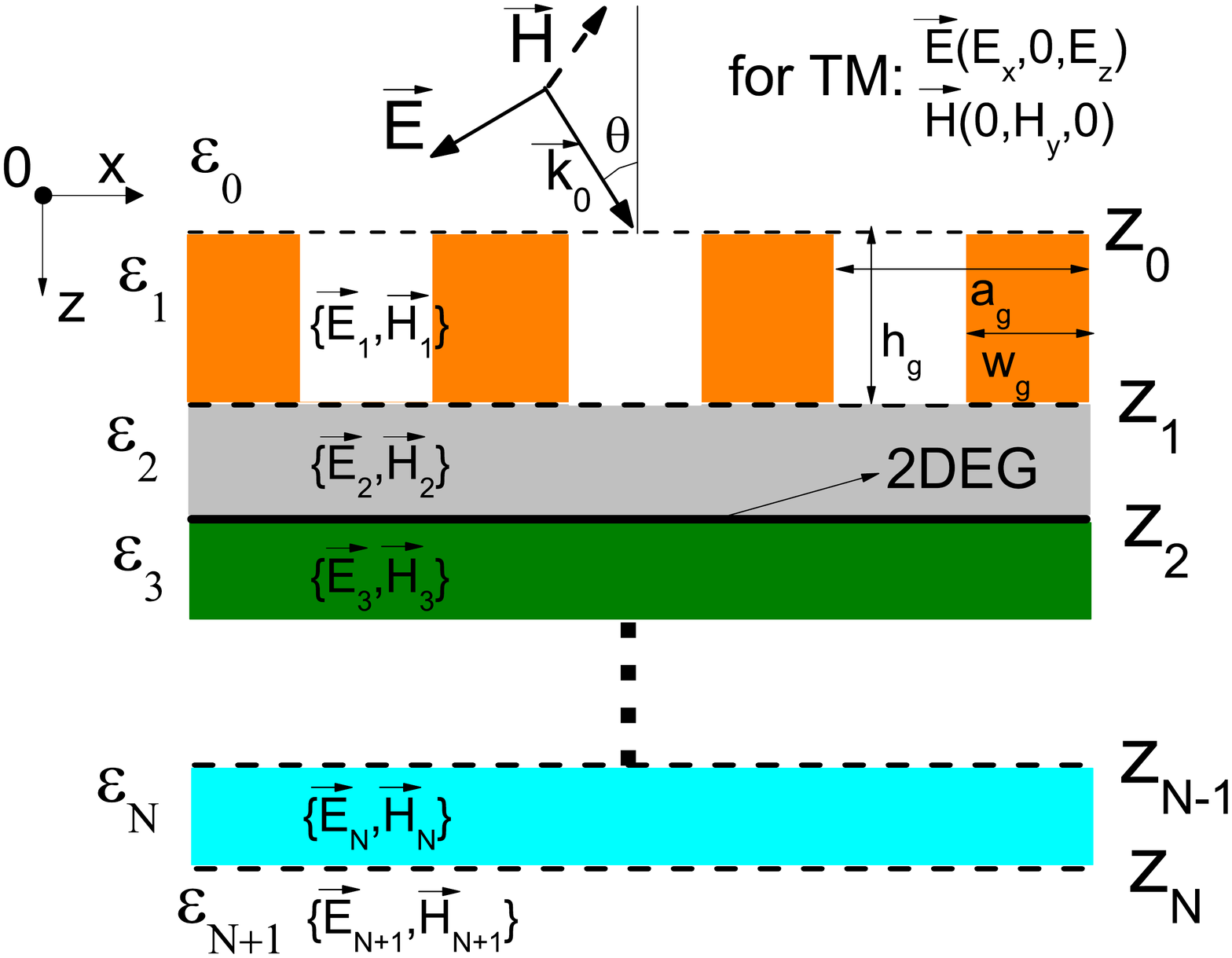}}
 \caption{A schematic sketch of the geometry of multilayered plasmonic structure with 2DEG. }
 \label{fig1}
\end{figure}

Let us assume that multilayered structure with the grating is illuminated by plane $em$ wave of TM polarization with frequency, $\omega$, and angle of incidence, $\theta$ (see Fig.\ref{fig1}). The grating of period $a_{g}$ is formed by the infinitely long in $y-$direction rectangular bars with width, $w_g$ and height, $h_{g}$. The structure consists of $N$ layers with thicknesses, $d_{j}=z_{j}-z_{j-1}$, $j=1...N$. Each $j-$layer, including the grating region, is described by the own dielectric permittivity, $\epsilon_{\omega,j}(x)$. The delta-thin conductive channel is placed between $j-1$ and $j$ layers and described by high-frequency conduction current, $\vec{J}^{2D}(x,t)\delta(z-z_{j})$. The whole structure is set between two non-absorbing half-spaces with $\epsilon_{0}$ (at $z<0$) and $\epsilon_{N+1}$
(at $z>D$, where $D$ is the total thickness of the structure).

Assuming that all components of $em$ field oscillate in time as $\exp(-i\omega t)$, the Maxwell's equations written for amplitudes take the form:
\begin{align}
\text{rot}\vec{H}_{\omega}&=-ik_{0}\epsilon_{\omega}(x,z)\vec{E}_{\omega}+\frac{4\pi}{c}\vec{J}^{2D}_{\omega}(x)\delta(z-z_{j}),\nonumber\\
\text{rot}\vec{E}_{\omega}&=ik_{0}\vec{H}_{\omega}. \label{Eq1}
\end{align}
where $k_{0}=\omega/c$, $\epsilon_{\omega}(x,z)=\epsilon_{\omega,j}(x)$ at $z\in[z_{j-1},z_{j}]$ and each non-zero components of the vectors $\vec{E}$ and $\vec{H}$ are the functions of $x$ and $z$ coordinates. For simplicity, we consider non-magnetic structure with magnetic permittivity is equal to $1$.
For the case of planar diffraction and TM polarization, the non-zero components are $E_{x},\,E_{z}$ and $H_{y}$.
Then, Eqs. (\ref{Eq1}) can be rewritten as:
\begin{align}
\frac{\partial H_{\omega,y}}{\partial z}&=ik_{0}\epsilon_{\omega}(x,z)E_{\omega,x}-\frac{4\pi}{c}J^{2D}_{\omega, x}(x)\delta(z-z_{j}),\nonumber\\
\frac{\partial H_{\omega,y}}{\partial x}&=-ik_{0}\epsilon_{\omega}(x,z)E_{\omega,z}, \nonumber\\
\frac{\partial E_{\omega,x}}{\partial z}&-\frac{\partial E_{\omega,z}}{\partial x}=ik_{0}H_{\omega,y}
\label{Eq2}
\end{align}
According the Floquet theorem, we can search for the solutions in the form of Fourier expansion:
\begin{equation}
\left(\!
\begin{array}{l}
H_{\omega,y}(x,z)  \\
E_{\omega,\{x,z\}}(x,z)  \\
J^{2D}_{\omega, x}(x)
\end{array}
\!\right)=\sum_{m=-M}^{M}\!\!
\left(\!
\begin{array}{l}
H_{\omega,m,y}(z)  \\
E_{\omega,m,\{x,z\}}(z)  \\
J^{2D}_{\omega,m,x}
\end{array}
\!\right)\!\exp(i\beta_{m}x),
\label{Eq3}
\end{equation}
where $\beta_{m}=k_{0}\sqrt{\epsilon_{0}}\sin\theta+q_{m}$ with $q_{m}=2\pi m/a_{g}$. The truncation rank $M$ (actual number of Fourier harmonics) is selected in such way to provide
the convergence of the solution with a given accuracy. Using the expansion (\ref{Eq3}), system (\ref{Eq2}) written for each $j-$layer in Fourier representation reads as:
\begin{align}
\frac{\partial {\bf H}_{y,j}}{\partial z}&=ik_{0} {\cal \hat{E}}^{-1}_{\text{inv},j}{\bf E}_{x,j},\nonumber\\
\hat{\beta}{\bf H}_{y,j} &=-k_{0} {\cal \hat{E}}_{j}{\bf E}_{z,j},\nonumber\\
\frac{\partial {\bf E}_{x,j}}{\partial z}&=ik_{0}{\bf H}_{y,j}+i \hat{\beta}{\bf E}_{z,j}.
\label{Eq4}
\end{align}
Here, we introduced the matrix notations: where Fourier-vectors ${\bf H}_{y,j}$, ${\bf E}_{\{x,z\},j}$ contain $2M+1$ corresponding Fourier components, $H_{\omega,m,y}$, $E_{\omega,m,\{x,z\}}$; $ \hat{\beta}$ is the diagonal matrix formed by the elements $\beta_{m}\delta_{m,m'}$ (here $\delta$ denotes the Kronecker delta symbol).
Elements of the matrix ${\cal \hat{E}}_{j}$ and ${\cal \hat{E}}_{\text{inv},j}$ are expressed through the spatial profile of the dielectric permittivity $\epsilon_{\omega,j}(x)$
\begin{align}
[{\cal \hat{E}}_{j}]_{m,m'}=\int_{0}^{1}\epsilon_{\omega,j}(\bar{x})\exp(-2\pi i[m-m']\bar{x})d\bar{x}, \nonumber\\
[{\cal \hat{E}}_{\text{inv},j}]_{m,m'}=\int_{0}^{1}\epsilon^{-1}_{\omega,j}(\bar{x})\exp(-2\pi i[m-m']\bar{x})d\bar{x},
 \label{Eq5}
\end{align}
respectively. Here $\bar{x}=x/a_{g}$.
Note, that the emergence of the ${\cal \hat{E}}^{-1}_{\text{inv},j}$ matrix in the first equation of system (\ref{Eq4}) relates to the Fourier factorization rule~\cite{Li, Popov2004} of a product of two discontinues functions $\epsilon_{\omega}(x)$ and $E_{\omega,x}(x)$ with concurrent jump in the region of the grating (product $\epsilon_{\omega}E_{\omega,x}$ is the electrical induction  which is continuous in the $x-$direction). In the spatially uniform layers, matrices ${\cal \hat{E}}^{-1}_{\text{inv},j}$ and ${\cal \hat{E}}_{j}$  are diagonal and identical.
The application of this rule considerably improves the convergence of the results comparing with the old formulation of RCWA method\cite{Moharam, Gaylord}, where classical Laurent rule, representing the product $\epsilon_{\omega} E_{\omega,x}$ as a convolution-type sum in conventional form   (${\cal \hat{E}}_{j}$ was used instead of ${\cal \hat{E}}^{-1}_{\text{inv},j}$).

System (\ref{Eq4}) can be rewritten in terms of the vector ${\bf H}_{y,j}$,
\begin{align}
\label{Eq6}
\frac{\partial^{2} {\bf H}_{y,j}}{\partial z^{2}}&=k_{0}^{2}{\bf \hat{A}}_{j}{\bf H}_{y,j},\quad \text{where}\\
{\bf \hat{A}}_{j}&={\cal \hat{E}}^{-1}_{\text{inv},j}\left[\frac{\hat{\beta}{\cal \hat{E}}^{-1}_{j}\hat{\beta}}{k_{0}^{2}}-\hat{I}\right],\nonumber
\end{align}
and $\hat{I}$ is the identity matrix. Having ${\bf H}_{y,j}$, we can find Fourier-vectors of the electric field components:
\begin{align}
 {\bf E}_{x,j}=-\frac{i}{k_{0}}{\cal \hat{E}}_{\text{inv},j}\frac{\partial {\bf H}_{y,j}}{\partial z},
 \quad {\bf E}_{z,j}=-\frac{1}{k_{0}}{\cal \hat{E}}^{-1}_{j}\hat{\beta}{\bf H}_{y,j}.
 \label{Eq7}
\end{align}
The system (\ref{Eq6}) should be solved consequently for each $j-$layer with appropriate boundary conditions on the $j-$interfaces:
 \begin{align}
 {\bf E}_{x,j}(z_{j})\!=\!{\bf E}_{x,j+1}(z_{j}),\, {\bf H}_{y,j}(z_{j})\!-\!{\bf H}_{y,j+1}(z_{j})\!=\!\frac{4\pi}{c}{\bf J}^{2D}_{x,j},
 \label{Eq8}
\end{align}
where ${\bf J}^{2D}_{x,j}$ is the Fourier-vector formed by $J^{2D}_{\omega,m,x}$ components. The first equation in (\ref{Eq8}) expresses the continuity of the tangential component of electric field and second equation describes the discontinuity of magnetic field component due to the presence of the conductive 2D layer~\cite{Blesz1993, Nosich1998}.
In the frames of the linear response theory, $J^{2D}_{\omega,m,x}=\sigma^{2D}_{\omega,m}E_{\omega,m,x}$, where high-frequency sheet conductivity, $\sigma^{2D}_{\omega,m}$, takes into account both frequency and spatial dispersion of the 2DEG. {Our approach is valid for arbitrary form of the conductivity of 2DEG. Particular examples of $\sigma^{2D}_{\omega,m}$ for the 2DEG with parabolic spectrum can be found in Refs.~\cite{Korot2018}, \cite{Michailov}. For the electrons with Dirac spectrum, as in the graphene, $\sigma^{2D}_{\omega,m}$  can be obtained using Kubo formalism (see Refs.~\cite{Gusynin2007, Falkovsky, Balaban2013}). In the case of doped graphene in steady-state applied electric field, $\sigma^{2D}_{\omega,m}$ can be found in Ref.~\cite{Kukhtaruk2016}.

Thus, the second boundary condition in Eqs.(\ref{Eq8}) can be rewritten in the  form:
  \begin{align}
 {\bf H}_{y,j}(z_{j})-\hat{\Gamma}^{2D}_{j}{\bf E}_{x,j}(z_{j})\!=\!{\bf H}_{y,j+1}(z_{j}),
 \label{Eq9}
\end{align}
where diagonal matrix $\hat{\Gamma}^{2D}_{j}$ is formed by the elements $4\pi/c\times\sigma^{2D}_{\omega,m}\delta_{m,m'}$.

The Eqs. (\ref{Eq6}) compose the system of ordinary second order differential equations with constant coefficients. The solution of this system can be expressed in terms of the eigen values, $\lambda_{m,j}$, and eigen vectors, $\vec{w}_{m,j}$ of the matrix ${\bf \hat{A}}$:
\begin{align}
 {\bf H}_{y,j}(z)\!=\!\sum_{\nu=1}^{2M+1}\!\vec{w}_{\nu,j}\left[C_{\nu,j}^{+}\exp(-k_{0}\bar{\lambda}_{\nu,j}(z\!-\!z_{j-1}))+ \right.\nonumber\\
 C_{\nu,j}^{-}\exp(k_{0}\bar{\lambda}_{\nu,j}(z\!-\!z_{j}))\left.\right],
 \label{Eq10}
\end{align}
where $\bar{\lambda}_{\nu,j}$ is the square root (with positive real part) of the eigenvalues $\lambda_{\nu,j}$. Two terms in the square brackets describe two waves: transmitted($+$) and reflected ($-$) into j-layer. The expression for Fourier vector, ${\bf E}_{x,j}(z)$ can be obtained by means
the first equation of (\ref{Eq7}) and written as follows
\begin{align}
 {\bf E}_{x,j}(z)\!=\!\sum_{\nu=1}^{2M+1}\!\vec{v}_{\nu,j}\left[C_{\nu,j}^{+}\exp(-k_{0}\bar{\lambda}_{\nu,j}(z\!-\!z_{j-1}))- \right.\nonumber\\
 C_{\nu,j}^{-}\exp(k_{0}\bar{\lambda}_{\nu,j}(z\!-\!z_{j}))\left.\right],
 \label{Eq11}
\end{align}
where vector $\vec{v}_{\nu,j}=i{\cal \hat{E}}_{\text{inv},j}\bar{\lambda}_{\nu,j}\vec{w}_{\nu,j}$.

Matching magnetic and electric fields on $j-$interface according to the boundary conditions (\ref{Eq8}) and (\ref{Eq9}), we come to the following recurrence relationship between constants of integration, $C_{\nu,j}^{\pm}$ and $C_{\nu,j+1}^{\pm}$:
\begin{align}
\sum_{\nu=1}^{2M+1}\vec{w}^{\,-}_{\nu,j}\exp(-k_{0}\bar{\lambda}_{\nu,j}d_{j})C_{\nu,j}^{+}+\vec{w}^{\,+}_{\nu,j}C_{\nu,j}^{-}&=\nonumber\\
\sum_{\nu=1}^{2M+1}\!\vec{w}_{\nu,j+1}\!\left[C_{\nu,j+1}^{+}\!+\!C_{\nu,j+1}^{-}\exp(-k_{0}\bar{\lambda}_{\nu,j+1}d_{j+1})\right];&\nonumber\\
\sum_{\nu=1}^{2M+1}\vec{v}_{\nu,j}\left[\exp(-k_{0}\bar{\lambda}_{\nu,j}d_{j})C_{\nu,j}^{+}-C_{\nu,j}^{-}\right]&=\nonumber\\
\sum_{\nu=1}^{2M+1}\!\vec{v}_{\nu,j+1}\!\left[C_{\nu,j+1}^{+}\!-\!C_{\nu,j+1}^{-}\exp(-k_{0}\bar{\lambda}_{\nu,j+1}d_{j+1})\right],
 \label{Eq12}
\end{align}
where $\vec{w}^{\,\pm}_{\nu,j}=\vec{w}_{\nu,j}\pm\hat{\Gamma}^{2D}_{j}\vec{v}_{\nu,j} $ contains parameters of 2DEG entered into matrix $\hat{\Gamma}^{2D}_{j}$.
The Eqs.(\ref{Eq12}) can be written in the compact matrix form:
\begin{align}
\!\!\left(\!\!\!\!
\begin{array}{ll}
{\bf \hat{W}}_{j}^{-} \hat{\Lambda}_{j}, &\!\! {\bf \hat{W}}_{j}^{+}\\
{\bf \hat{V}}_{j} \hat{\Lambda}_{j}, & \!\!-{\bf \hat{V}}_{j}
\end{array}
\!\!\!\!\right)
\!\!\!\left(\!\!
\begin{array}{l}
\vec{C}_{j}^{+}\\
\vec{C}_{j}^{-}\\
\end{array}
\!\!\right)\!\!\!=
\!\!\!\left(\!\!\!\!
\begin{array}{ll}
{\bf \hat{W}}_{j+1}, &\!\! {\bf \hat{W}}_{j+1}\hat{\Lambda}_{j+1}\\
{\bf \hat{V}}_{j+1}, &\!\! -{\bf \hat{V}}_{j+1}\hat{\Lambda}_{j+1}
\end{array}
\!\!\!\!\right)\!\!
\left(\!\!
\begin{array}{l}
\vec{C}_{j+1}^{+}\\
\vec{C}_{j+1}^{-}\\
\end{array}
\!\!\right)
\label{Eq13}
\end{align}
where each vector, $\vec{C}_{j}^{\pm}$, contains $2M+1$ integration constants, the matrixes ${\bf \hat{W}}_{j}$ and ${\bf \hat{V}}_{j}$ are formed by elements of the corresponding eigen vectors, $\vec{w}_{\nu,j}$ and  $\vec{v}_{\nu,j}$,  defined for all eigen values, $\lambda_{\nu,j}$. Matrixes
${\bf \hat{W}}_{j}^{\pm}={\bf \hat{W}}_{j}\pm \hat{\Gamma}^{2D}_{j}{\bf \hat{V}}_{j}$, $\hat{\Lambda}_{j}$ is the diagonal matrix with the elements,
$\exp(-k_{0}\bar{\lambda}_{\nu,j}d_{j})\delta_{\nu,\nu'}$.

Relationship (\ref{Eq13}) allows us to couple amplitudes of the reflected wave (in region $z<z_{0}$) and transmitted wave (in the region $z>z_{N}$), and consequently to calculate transmission and reflection coefficients for different diffraction orders. Indeed, components of the Fourier- vectors of
$y-$magnetic and $x-$electric fields in the region $z<0$ ($j=0$) are:
\begin{align}
\label{Eq14}
H_{\omega,m,y}&\!=\!\delta_{m,0}\exp(-\bar{\lambda}_{m,0}k_{0}z)+r_{\omega,m}\exp(\bar{\lambda}_{m,0}k_{0}z),\\
E_{\omega,m,x}&\!=\!\frac{\cos\theta}{\sqrt{\epsilon_{0}}}\delta_{m,0}\exp(-\bar{\lambda}_{m,0}k_{0}z)\!-\!
\frac{i\bar{\lambda}_{m,0}}{\epsilon_{0}}r_{\omega,m}\exp(\bar{\lambda}_{m,0}k_{0}z),
\nonumber
\end{align}
and in the region $z>D$ ($j=N+1$) are:
\begin{align}
\label{Eq15}
H_{\omega,m,y}&=t_{\omega,m}\exp(-\bar{\lambda}_{m,N+1}k_{0}(z-z_{N})),\\
E_{\omega,m,x}&=\frac{i\bar{\lambda}_{m,N+1}}{\epsilon_{N+1}}t_{\omega,m}\exp(-\bar{\lambda}_{m,N+1}k_{0}(z-z_{N})),
\nonumber
\end{align}
where $\bar{\lambda}_{m,\{0,N+1\}}=\sqrt{\beta_{m}^2-\epsilon_{\{0,N+1\}}k_{0}^2}/k_{0}$ if $\beta_{m}>\sqrt{\epsilon_{\{0,N+1\}}}k_{0}$ and
$\bar{\lambda}_{m,\{0,N+1\}}=-i\sqrt{\epsilon_{\{0,N+1\}}k_{0}^2-\beta_{m}^2}/k_{0}$ if otherwise. The quantities $r_{\omega,m}$ and $t_{\omega,m}$ are the normalized magnetic-field amplitudes
of the m-th backward-diffracted (reflected) and forward diffracted (transmitted) waves, respectively. These amplitudes form the Fourier-vectors ${\bf R}$ and ${\bf T}$.

Using Eqs. (\ref{Eq14}), (\ref{Eq15}), boundary conditions (\ref{Eq8}) and (\ref{Eq9}), and relationship (\ref{Eq13}), we found that Fourier-vectors ${\bf R}$ and ${\bf T}$ are coupled through following matrix equations:
\begin{align}
\!\!\left(\!\!\!
\begin{array}{c}
{\bf \hat{I}}-\frac{\cos\theta}{\sqrt{\epsilon_{0}}}\hat{\Gamma}^{2D}_{0} \\
\frac{\cos\theta}{\sqrt{\epsilon_{0}}}{\bf \hat{I}}
\end{array}
\!\!\right)\vec{\delta}+
\!\!\left(\!
\begin{array}{c}
{\bf \hat{I}}-\hat{\bf Z}_{I}\hat{\Gamma}^{2D}_{0} \\
\hat{\bf Z}_{I}
\end{array}
\right){\bf R}=\nonumber \\
\prod_{j=1}^{N}
\left(\!
\begin{array}{cc}
{\bf \hat{W}}_{j}, &\!\! {\bf \hat{W}}_{j}\hat{\Lambda}_{j}\\
{\bf \hat{V}}_{j}, &\!\! -{\bf \hat{V}}_{j}\hat{\Lambda}_{j}
\end{array}
\right)
\!\left(\!
\begin{array}{cc}
{\bf \hat{W}}_{j}^{-} \hat{\Lambda}_{j}, &\!\! {\bf \hat{W}}_{j}^{+}\\
{\bf \hat{V}}_{j} \hat{\Lambda}_{j}, & \!\!-{\bf \hat{V}}_{j}
\end{array}
\right)^{\!-1}
\!\left(\!
\begin{array}{c}
{\bf \hat{I}}\\
\hat{\bf Z}_{II}
\end{array}
\right){\bf T}\!,
\label{Eq16}
\end{align}
where matrices $\hat{\bf Z}_{I}$ and $\hat{\bf Z}_{II}$ are diagonal with elements $-i\bar{\lambda}_{m,0}/\epsilon_{0}\delta_{m,m'}$ and $i\bar{\lambda}_{m,N+1}/\epsilon_{N+1}\delta_{m,m'}$, respectively,
 $\vec{\delta}$ is the vector with elements ${\delta}_{m,0}$.
The (\ref{Eq16}) is the master system of equations of the modified RCWA method providing the way for calculation of transmission and reflection coefficients for different diffraction orders.
In contrast to the previous formulation of the RCWA method \cite{Moharam1995,Gaylord}, account of 2DEG leads to the nontrivial modifications. Particularly, the master system (\ref{Eq16}) contains the matrices ${\bf \hat{W}}_{j}^{\pm}$ in the right-hand side and $\hat{\Gamma}^{2D}_{0}$ in the left-hand side. Latter term describes the possible existence of the 2DEG on the top of the grating.

It should be noted that usage of this system of equations, written in the present form, applying to the situation of deep surface grating and optically dense materials
can face with problem of the computational instability. This instability is associated with the procedure of the numerical inversion of the second matrix in the r-h-s of the (\ref{Eq16})
when exponentially small terms, $\exp(-k_{0}\bar{\lambda}_{\nu,j}d_{j})$ (standing in the $\hat{\Lambda}_{j}$) becomes smaller than machine precision. To avoid this obstacle, authors in Ref.\cite{Gaylord} proposed to
use the following  decomposition of the badly inverted matrix:
\begin{align}
\left(\!\!\!
\begin{array}{cc}
{\bf \hat{W}}_{j}^{-} \hat{\Lambda}_{j}, &\!\! {\bf \hat{W}}_{j}^{+}\\
{\bf \hat{V}}_{j} \hat{\Lambda}_{j}, & \!\!-{\bf \hat{V}}_{j}
\end{array}
\!\!\!\right)^{\!\!-1}\!\!=\!\left(\!
\begin{array}{cc}
\hat{\Lambda}_{j}, & {\bf \hat{0}}\\
{\bf \hat{0}}, & {\bf \hat{I}}
\end{array}
\right)^{\!\!-1}
\!\!\left(\!\!
\begin{array}{cc}
{\bf \hat{W}}_{j}^{-}, & {\bf \hat{W}}_{j}^{+} \\
{\bf \hat{V}}_{j}, & -{\bf \hat{V}}_{j}
\end{array}
\!\!\right)^{\!\!-1}
\label{Eq17}
\end{align}
Note, that second inversion matrix in the r-h-s of Eq.~(\ref{Eq17}) is the regular and can be numerically inverted without any difficulties.
Let's the product of the last $N-$terms in Eq.~(\ref{Eq16})
\begin{align}
\left(\!\!\!
\begin{array}{cc}
{\bf \hat{W}}_{N}, &\!\! {\bf \hat{W}}_{N}\hat{\Lambda}_{N}\\
{\bf \hat{V}}_{N}, &\!\! -{\bf \hat{V}}_{N}\hat{\Lambda}_{N}
\end{array}
\!\!\!\right)
\left(\!\!\!\!
\begin{array}{cc}
\hat{\Lambda}_{N}, & {\bf \hat{0}}\\
{\bf \hat{0}}, & {\bf \hat{I}}
\end{array}
\!\!\!\!\right)^{\!\!-1}
\left(\!\!
\begin{array}{c}
{\bf \hat{X}}_{N}\\
{\bf \hat{Y}}_{N}
\end{array}
\!\!\right)\!\!{\bf T}\equiv\left(\!\!\!
\begin{array}{c}
{\bf \hat{f}}_{N}\\
\hat{\bf g}_{N}
\end{array}
\!\!\!\right){\bf T}
\end{align}
where we introduced the following designation
\begin{align}
\left(
\begin{array}{c}
{\bf \hat{X}}_{N}\\
{\bf \hat{Y}}_{N}
\end{array}
\right)\equiv
\left(\!
\begin{array}{cc}
{\bf \hat{W}}_{N}^{-}, &\!\! {\bf \hat{W}}_{N}^{+}\\
{\bf \hat{V}}_{N}, & \!\!-{\bf \hat{V}}_{N}
\end{array}
\right)^{\!\!-1}
\left(
\begin{array}{c}
{\bf \hat{I}}\\
\hat{\bf Z}_{II}
\end{array}
\right).
\nonumber
\end{align}
Making substitution, ${\bf T}={\bf {\hat X}}_{N}^{-1}\hat{\Lambda}_{N}{\bf T}_{N}$, term
\begin{align}
\left(\!
\begin{array}{cc}
\hat{\Lambda}_{N}, & {\bf \hat{0}}\\
{\bf \hat{0}}, & {\bf \hat{I}}
\end{array}
\right)^{\!\!-1}\!\!
\left(
\begin{array}{c}
{\bf \hat{X}}_{N}\\
{\bf \hat{Y}}_{N}
\end{array}
\right){\bf T}\!=\!\left(\!
\begin{array}{cc}
\hat{\Lambda}_{N}, & {\bf \hat{0}}\\
{\bf \hat{0}}, & {\bf \hat{I}}
\end{array}
\right)^{\!\!-1}
\!\!\left(
\begin{array}{c}
\hat{\Lambda}_{N}\\
{\bf \hat{Y}}_{N}{\bf \hat{X}}_{N}^{-1}\hat{\Lambda}_{N}
\end{array}
\right)&\nonumber\\
\times{\bf T}_{N}=\left(\!
\begin{array}{cc}
\hat{\Lambda}_{N}, & {\bf \hat{0}}\\
{\bf \hat{0}}, & {\bf \hat{I}}
\end{array}
\right)^{\!\!-1}\!\!
\left(\!
\begin{array}{cc}
\hat{\Lambda}_{N}, & {\bf \hat{0}}\\
{\bf \hat{0}}, & {\bf \hat{I}}
\end{array}
\right)\!\!
\left(
\begin{array}{c}
{\bf \hat{I}}\\
{\bf \hat{Y}}_{N}{\bf \hat{X}}_{N}^{-1}\hat{\Lambda}_{N}
\end{array}
\right){\bf T}_{N} & \nonumber\\
=\left(
\begin{array}{c}
{\bf \hat{I}}\\
{\bf \hat{Y}}_{N}{\bf \hat{X}}_{N}^{-1}\hat{\Lambda}_{N}
\end{array}
\right){\bf T}_{N}\!, &\nonumber
\end{align}
and we can obtain that
\begin{align}
\left(
\begin{array}{c}
{\bf \hat{f}}_{N}\\
\hat{\bf g}_{N}
\end{array}
\right){\bf T}=
\left(\begin{array}{c}
{\bf \hat{W}}_{N}\left[{\bf \hat{I}}+\hat{\Lambda}_{N}{\bf \hat{Y}}_{N}{\bf \hat{X}}_{N}^{-1}\hat{\Lambda}_{N}\right]\\
{\bf \hat{V}}_{N}\left[{\bf \hat{I}}-\hat{\Lambda}_{N}{\bf \hat{Y}}_{N}{\bf \hat{X}}_{N}^{-1}\hat{\Lambda}_{N}\right]
\end{array}
\right){\bf T}_{N}
\label{Eq19}
\end{align}
Substituting relationship (\ref{Eq19}) into Eqs.(\ref{Eq16}) and sequentially performing above mentioned transformations for each $j$-th term in the product
we can rewrite master system of the equations (\ref{Eq16}) in the computationally stable form:
\begin{align}
\!\!\left(\!\!\!
\begin{array}{c}
{\bf \hat{I}}-\frac{\cos\theta}{\sqrt{\epsilon_{0}}}\hat{\Gamma}^{2D}_{0} \\
\frac{\cos\theta}{\sqrt{\epsilon_{0}}}{\bf \hat{I}}
\end{array}
\!\!\!\right)\vec{\delta}+
\!\!\left(\!\!\!
\begin{array}{c}
{\bf \hat{I}}-\frac{\cos\theta}{\sqrt{\epsilon_{0}}}\hat{\Gamma}^{2D}_{0} \\
\hat{\bf Z}_{I}
\end{array}
\!\!\!\right)\!{\bf R}=
\!\left(\!
\begin{array}{c}
{\bf \hat{f}}_{1}\\
{\bf \hat{g}}_{1}
\end{array}
\!\right)\!{\bf T}_{1}.
\label{Eq20}
\end{align}
The matrices ${\bf \hat{f}}_{1}$ and ${\bf \hat{g}}_{1}$ can be found from the following recurrence relationship:
\begin{align}
\!\left(\!\!\!
\begin{array}{c}
{\bf \hat{f}}_{j-1}\\
{\bf \hat{g}}_{j-1}
\end{array}
\!\!\!\right)=
\left(\!\!\!
\begin{array}{c}
{\bf \hat{W}}_{j-1}\left[{\bf \hat{I}}+\hat{\Lambda}_{j-1}{\bf \hat{Y}}_{j-1}{\bf \hat{X}}_{j-1}^{-1}\hat{\Lambda}_{j-1}\right]\\
{\bf \hat{V}}_{j-1}\left[{\bf \hat{I}}-\hat{\Lambda}_{j-1}{\bf \hat{Y}}_{j-1}{\bf \hat{X}}_{j-1}^{-1}\hat{\Lambda}_{j-1}\right]
\end{array}
\!\!\!\right),
\label{Eq21}
\end{align}
where
\begin{align}
\!\left(\!
\begin{array}{c}
{\bf \hat{X}}_{j-1}\\
{\bf \hat{Y}}_{j-1}
\end{array}
\!\right)=
\left(\!
\begin{array}{cc}
{\bf \hat{W}}_{j-1}^{-}, &\!\! {\bf \hat{W}}_{j-1}^{+}\\
{\bf \hat{V}}_{j-1}, & \!\!-{\bf \hat{V}}_{j-1}
\end{array}
\!\right)^{-1}
\!\left(\!
\begin{array}{c}
{\bf \hat{f}}_{j}\\
{\bf \hat{g}}_{j}
\end{array}
\!\right)
\label{Eq22}
\end{align}
and $j$ is varied from $N$ to $2$.
The vector ${\bf T}_{1}$ relates to Fourier-vector of transmission coefficient, ${\bf T}$, as follows
\begin{align}
{\bf T}=\prod_{j=N}^{1}{\bf \hat{X}}_{j}^{-1}\hat{\Lambda}_{j}{\bf T}_{1}.
\label{Eq23}
\end{align}

Components of the vectors ${\bf T}$ and ${\bf R}$  provide the transmission, ${\it T}_{m}$, reflection, ${\it R}_{m}$, coefficients for any m-th diffraction order as well as total absorption, ${\it L}$. Particularly,
\begin{align}
T_{m}&=\frac{\sqrt{\epsilon_{0}}\sqrt{\epsilon_{N+1}k_{0}^2-\beta_{m}^2}}{k_{0}\epsilon_{N+1}\cos\theta}|{\bf T}[m]|^2,\nonumber\\
R_{m}&=\frac{\sqrt{\epsilon_{0}k_{0}^2-\beta_{m}^2}}{k_{0}\sqrt{\epsilon_{0}}\cos\theta}|{\bf R}[m]|^2,
L=1-\sum_{m}{\it T}_{m}+{\it R}_{m}.
\label{Eq24}
\end{align}
Here, the summation is taken over all numbers of visible diffraction orders, i.e, over such values of $m$ which keep the positive expressions under the square roots in the nominators. In the case of subwavelength gratings, only zero diffraction order ($m=0$) occurs and we have that
 $T_{0}=|{\bf T}[0]|^2$, $R_{0}=|{\bf R}[0]|^2$ (if  $\epsilon_{0}=\epsilon_{N+1}$), and $L\equiv L_{0}=1-T_{0}-R_{0}$.
The Eqs.~(\ref{Eq20}-\ref{Eq23}) together with Eqs.~(\ref{Eq24}) finalize the computationally stable realization of the modified RCWA method for characterization of the multi-layered plasmonic
structures with grating-gated delta-thin conductive channels in the case of TM-polarization.

This method can be extended for the case of TE polarization of the incident radiation. Now, actual components of the $em$ waves are: $E_{y}$, $H_{x}$ and $H_{z}$. Matrix equation (\ref{Eq6})
is formulated for the Fourier-vectors ${\bf E}_{y,j}$, where
\begin{equation}
{\bf \hat{A}}_{j}=\hat{\beta}^{2}/k_{0}^{2}- {\cal \hat{E}}_{j}.
\label{Eq25}
\end{equation}
The Fourier-vectors of the magnetic field components are given as follows:
$$   \mathbf{H}_{x, j}  = \frac{i}{k_0} \frac{\partial \mathbf{E}_{y, j}}{\partial z},\,\,   \mathbf{H}_{z, j}  =  \frac{\hat{\beta}}{k_0} \mathbf{E}_{y, j}$$  and boundary conditions reads as
$$\mathbf{E}_{y, j}(z_j)\! =\! \mathbf{E}_{y, j+1}(z_j),\,\mathbf{H}_{x, j}(z_j) + \hat{\Gamma}^{2D}_j \mathbf{E}_{y, j}(z_j)\!=\!\mathbf{H}_{x, j+1}(z_{j})\,.$$ Finally, master system of equations (\ref{Eq16}) takes the form:
\begin{align}
\!\!\left(\!
\begin{array}{c}
{\bf \hat{I}} \\
\hat{\Gamma}^{2D}_{0}-\sqrt{\epsilon_{0}}\cos\theta{\bf \hat{I}}
\end{array}
\right)\vec{\delta}+
\!\!\left(\!
\begin{array}{c}
{\bf \hat{I}} \\
\hat{\Gamma}^{2D}_{0}+\hat{\bf Z}_{I}
\end{array}
\right){\bf R}=\nonumber \\
\prod_{j=1}^{N}
\left(\!
\begin{array}{cc}
{\bf \hat{W}}_{j}, &\!\! {\bf \hat{W}}_{j}\hat{\Lambda}_{j}\\
{\bf \hat{V}}_{j}, &\!\! {\bf \hat{V}}_{j}\hat{\Lambda}_{j}
\end{array}
\right)
\!\left(\!
\begin{array}{cc}
{\bf \hat{W}}_{j} \hat{\Lambda}_{j}, &\!\! {\bf \hat{W}}_{j}\\
{\bf \hat{V}}_{j}^{+} \hat{\Lambda}_{j}, & \!\!{\bf \hat{V}}_{j}^{-}
\end{array}
\right)^{\!-1}
\!\left(\!
\begin{array}{c}
{\bf \hat{I}}\\
\hat{\bf Z}_{II}
\end{array}
\right){\bf T}\!,
\label{Eq26}
\end{align}
where ${\bf \hat{W}}_{j}$ is formed by eigen vectors,  $\vec{w}_{\nu,j}$ obtained for the each $\lambda_{\nu,j}$ eigen values of the matrix (\ref{Eq25}). Matrix ${\bf \hat{V}}_{j}$ contains vectors, $\vec{v}_{\nu,j}=-i\bar{\lambda}_{\nu,j}\vec{w}_{\nu,j}$ and ${\bf \hat{V}}_{j}^{\pm}={\bf \hat{V}}_{j}\pm \hat{\Gamma}^{2D}_{j}{\bf \hat{W}}_{j}$. The matrices $\hat{\bf Z}_{I}$ and $\hat{\bf Z}_{II}$ are
the diagonal with elements $i\bar{\lambda}_{m,0}\delta_{m,m'}$ and $-i\bar{\lambda}_{m,N+1}\delta_{m,m'}$, respectively. Applying procedure (see above) for stable computation of system of equations (\ref{Eq26}), we can find vectors ${\bf T}$ and ${\bf R}$, and calculate transmission, reflection coefficients for any m-th diffraction order as well as total absorption for the case of TE-polarized incident radiation:
\begin{align}
T_{m}&=\frac{\sqrt{\epsilon_{N+1}k_{0}^2-\beta_{m}^2}}{k_{0}\sqrt{\epsilon_{0}}\cos\theta}|{\bf T}[m]|^2,
R_{m}=\frac{\sqrt{\epsilon_{0}k_{0}^2-\beta_{m}^2}}{k_{0}\sqrt{\epsilon_{0}}\cos\theta}|{\bf R}[m]|^2,\nonumber\\
L&=1-\sum_{m}{\it T}_{m}+{\it R}_{m}.
\label{Eq27}
\end{align}

The proposed modified RCWA method is applied for investigation of particular plasmonic structures with grating-gated 2DEG channel. Such structures possess the resonant properties in the THz frequency range (wavelength of order of 100 \textmu m) for TM-polarized incident radiation due to excitations of plasmons in conductive channel of 2DEG~\cite{Michailov, Popov, Korot2014, Pashnev2020}.   Below,  we will study spectral characteristics  of the  AlGaN/GaN-based plasmonic structure with deeply subwavelength (micron period) metallic grating, calculating transmission ($T_{0}$), reflection ($R_{0}$) and absorption ($L_{0}$) coefficients, their convergence vs number of the Fourier harmonics and the near-field mapping. Also, we will pay attention to the dependence of the plasmon resonances vs geometry of the grating.

\section{Far-field characteristics and their convergence}\label{Sec2}
Here and below, we will study the case of TM-polarized incident $em$ wave with incidence angle, $\theta=0$.
The structure under test is formed by $N=3$ media, embedded into the air, including the rectangular metallic grating with $\epsilon_{\omega,1}(x)=\epsilon_{\omega, M}\Theta(w_{g}-x)+\epsilon_{0}\Theta(x-w_{g})$
(where $\Theta(x)$ stands the Heaviside step function, $x\in [0,a_{g}]$,  $\epsilon_{\omega, M}=1+4\pi i\sigma_{M}/\omega$ and $\sigma_{M}=4\times 10^{17}$ s$^{-1}$ that corresponds to the gold),
AlGaN barrier and GaN buffer layers with constant dielectric permittivities $\epsilon_{2}=9.2$, and $\epsilon_{3}=8.9$, respectively. The 2D conductive channel is formed in the plane $z=z_{2}$.
The matrix $[\Gamma^{2D}_{j}]_{m,m'}=4\pi/c\times\sigma^{2D}_{\omega,m}\delta_{m,m'}\delta_{j,2}$, where for description of the high-frequency properties of 2DEG we used Drude-Lorentz
model $\sigma^{2D}_{\omega,m}=e^{2}n_{2D}\tau_{2D}/m^{*}(1-i\omega\tau_{2D})$ with electron effective mass, $m^{*}=0.22\times m_{e}$, concentration of 2DEG, $n_{2D}=6\times 10^{12}$~cm$^{-2}$ and effective scattering time, $\tau_{2D}=0.5$~ps. Other geometrical parameters of the structure are listed in the caption to Fig.\ref{fig2}. The selected parameters are close to the parameters of the experimental structures recently studied in Ref.\cite{Pashnev2020}.
\begin{figure}[t!]
\centering
\includegraphics[width=0.4\textwidth]{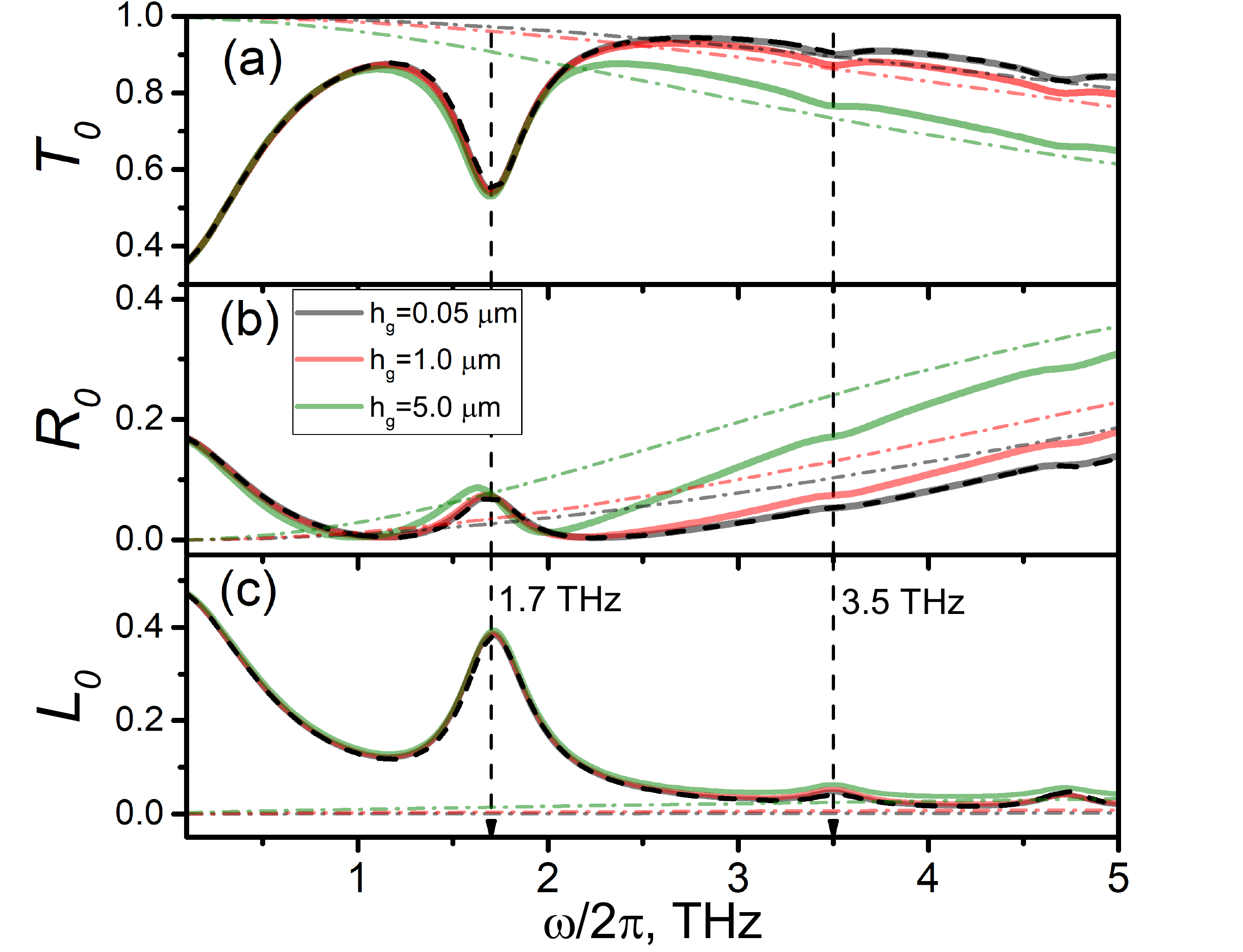}
\includegraphics[width=0.4\textwidth]{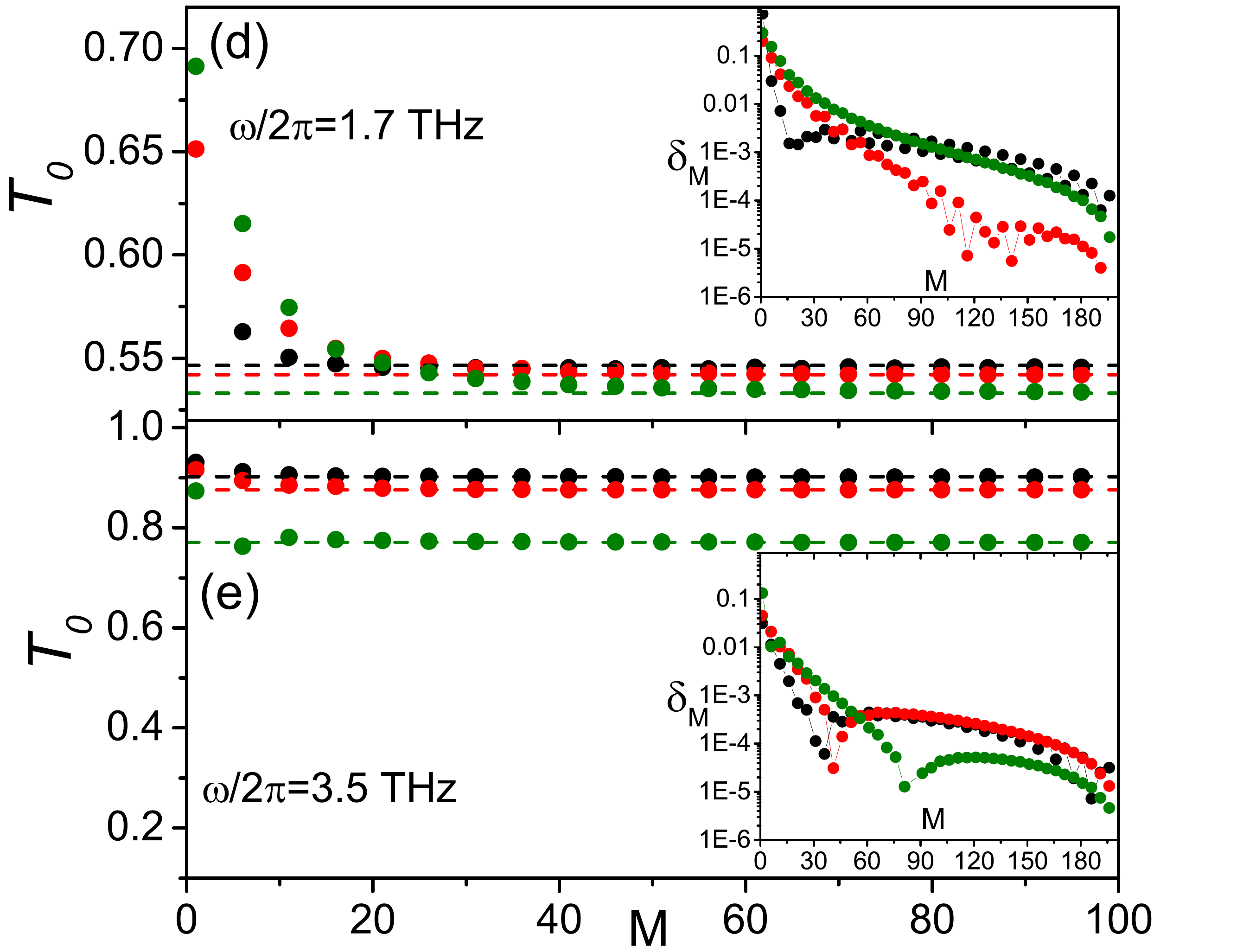}
 \caption{(color online): Spectra of the transmission (a), reflection (b) and absorption (c) coefficients for zero diffraction order calculated at three values of the grating depth: $h_{g}\equiv d_{1}=0.05,\,1,\, 5$ \textmu m.
 Grating period, $a_{g}=1$ \textmu m and width of grating bars, $w_{g}=0.5$ \textmu m. Thickness of AlGaN barrier, $d_{2}=0.025$  \textmu m and GaN buffer layer, $d_{3}=1$  \textmu m.
 Black dashed lines are the results of the IE method assuming delta-thin grating with 2D conductivity, $\sigma^{2D}_{g}=2\times 10^{12}$ cm/s.
 Dash-dotted lines are the results for the structure without 2DEG. All spectra are obtained at $M=100$. Panels (d) and (e): Dependencies of the transmission coefficient, ${\it T}_{0}$ vs number of Fourier harmonics, $M$,
 for two selected frequencies. Insets: relative errors, $\delta_{M}$, are plotted in logarithmic scale as function of $M$. }
 \label{fig2}
\end{figure}

The spectra of the far-field characteristics such as transmission, reflection and absorption coefficients calculated for three depths of the metallic grating are illustrated in Fig.\ref{fig2}. As seen, all spectra possess a strong resonance at the frequency of $1.7$ THz and much more weaker resonances at frequencies of $3.5$ THz and $4.75$ THz. The emergence of these resonances relate to the grating-assisted interaction of incident $em$ wave with the plasmons in the channel of the 2DEG. At the resonance, plasmon excitation of 2DEG with wavevectors determined by the grating period can effectively absorb energy of the incident $em$ wave.
Physics of 2D plasmons and their resonant interaction with $em$ radiation are well-described in Refs.~\cite{Popov, Michailov, Korot2014, Chaplik, DyakonovPRL93}.
\begin{figure}[t]
\centering
\includegraphics[width=0.4\textwidth]{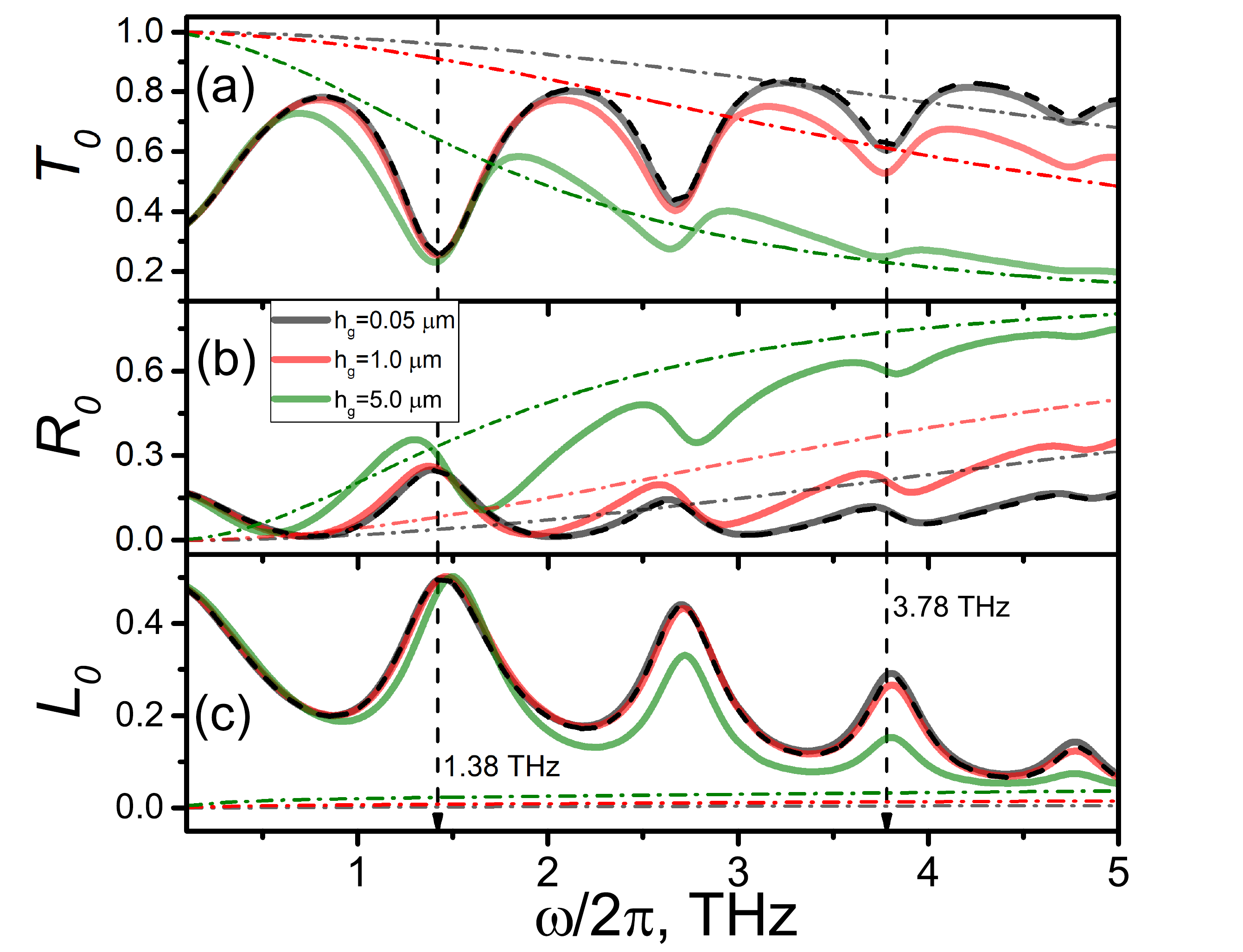}
\includegraphics[width=0.4\textwidth]{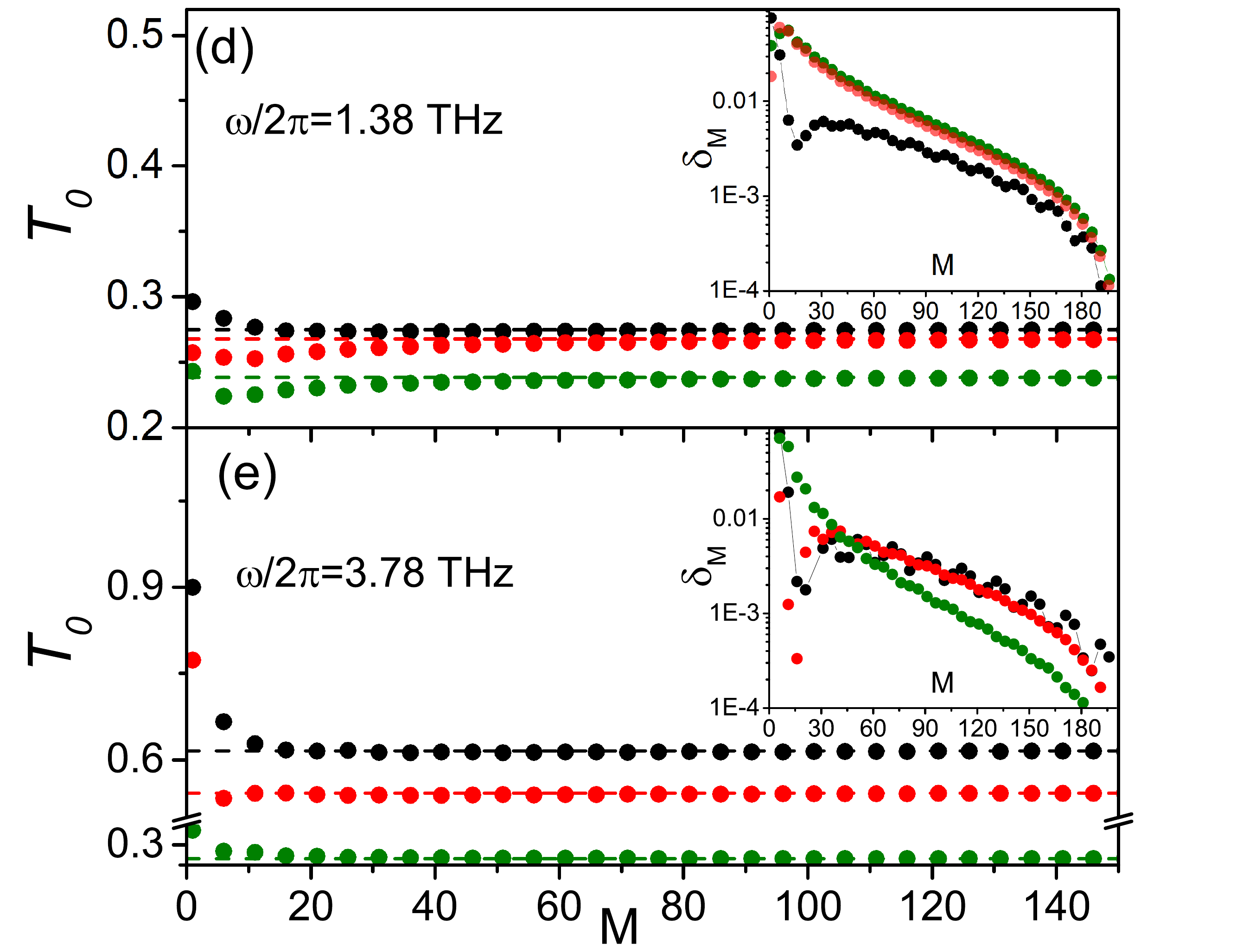}
 \caption{(color online): The same as in Fig.\ref{fig2} at $w_{g}=0.85$ \textmu m.}
 \label{fig3}
\end{figure}

The considered plasmonic structure can provide considerable absorption of the THz radiation. For example, at 1.7 THz the absorption coefficient $L_{0}\sim 40\%$. The RCWA-calculations show that absorption coefficient of the plasmonic structure is almost independent on the depth of the metallic grating (see Fig.\ref{fig2}(c)). Moreover, we show that results of IE method~\cite{Michailov, Popov2010, Korot2017}, developed for the same plasmonic structure but delta-thin grating (see black dashed lines), and  present RCWA for the grating with $h_{g}=0.05$ \textmu m almost coincide. An increase of the
grating depth only leads to the additional dispersion of transmission and reflection coefficients in the higher frequency range. This dispersion is also observed in modeling structure without 2DEG (see dashed-dotted lines).
Very weak dependence of the absorption of the plasmonic structure vs grating depth (even for deep grating with $h_{g}=5$ \textmu m) indicates that subwavelength highly conductive grating plays the role of almost-lossless waveguide for incident TM-polarized wave of THz frequencies. For example, absorption of the deep grating in the structure without 2DEG does not exceed $2\%$ in the considered spectral range.
The calculations of the partial losses associated with grating and 2DEG gas in the grating-2DEG plasmonic structure can be performed using the pattern of the near-field and one will be done below in the Sec.\ref{Sec3}.

\begin{figure*}
\centering
\includegraphics[width=0.32\textwidth]{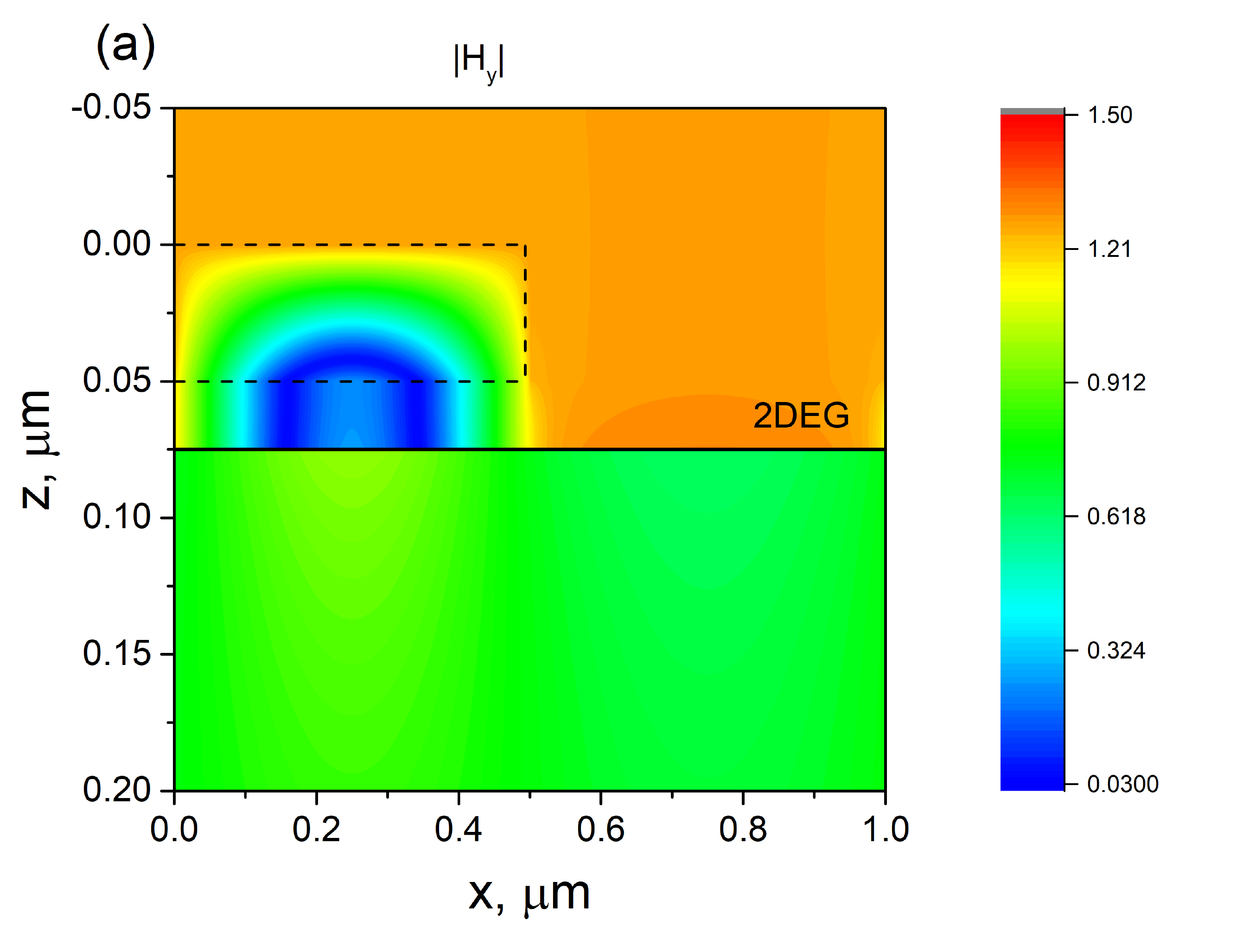}
\includegraphics[width=0.32\textwidth]{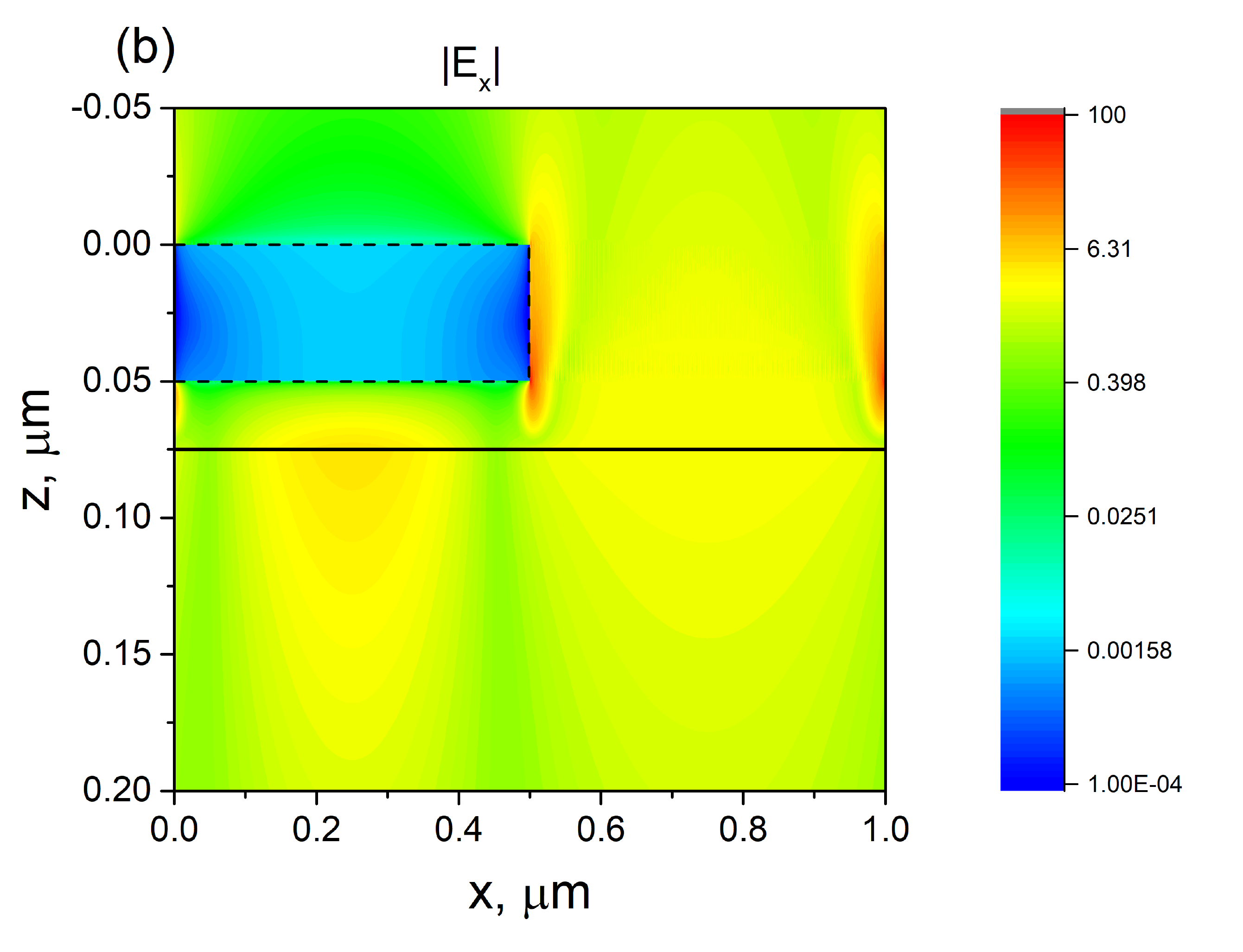}
\includegraphics[width=0.32\textwidth]{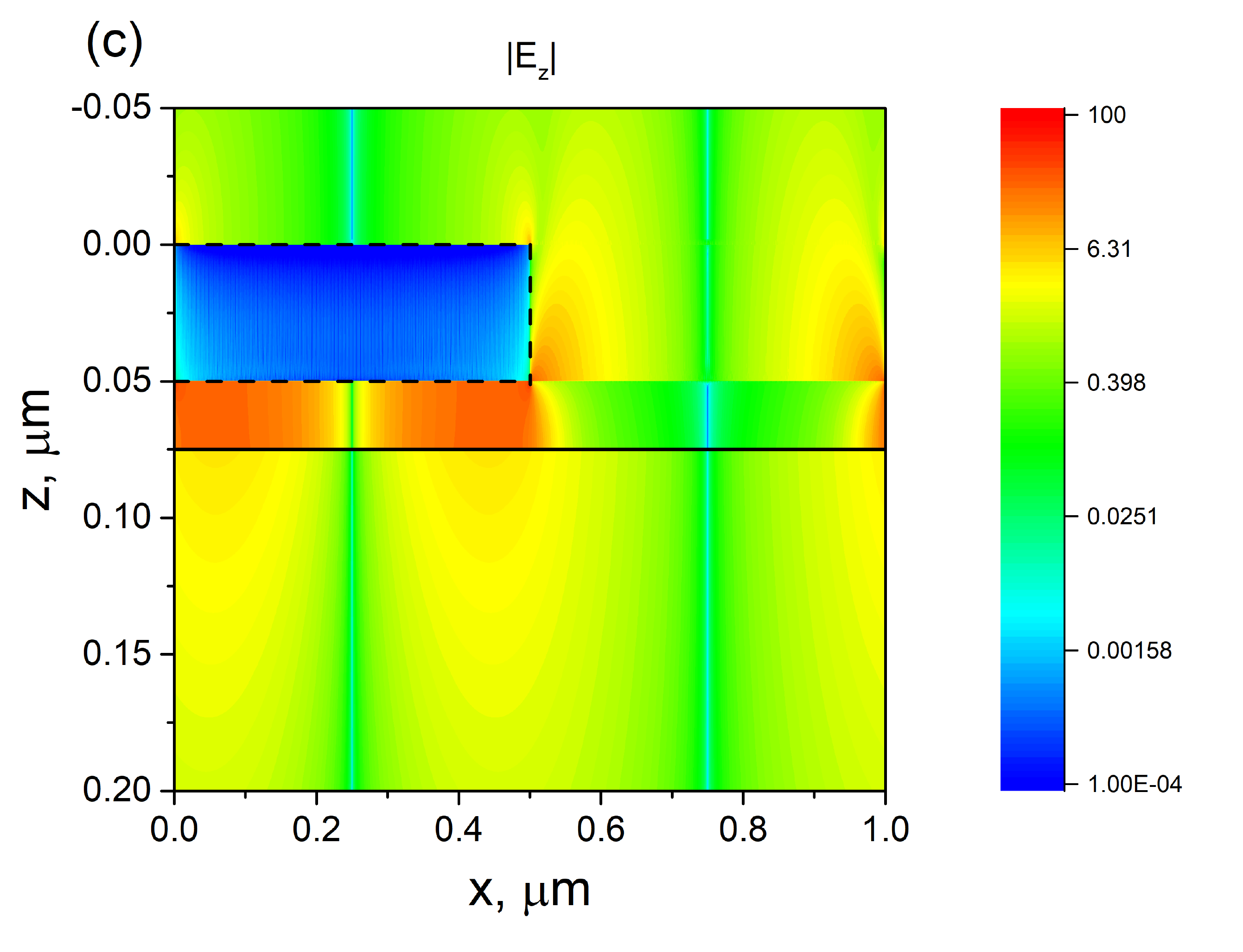}
\caption{Spatial distributions of amplitudes of $H_{y}$ (a), $E_{x}$ (b) and $E_{z}$ (c) components of $em$ field in the units of amplitude of incident wave, $E_{ins}$, for the structure with $a_{g}=1$ \textmu m, $w_{g}=0.5$ \textmu m, $h_{g}=0.05 $ \textmu m at frequency $\omega/2\pi=1.7$ THz. Number of Fourier harmonics, $M=450$.}
\label{fig4}
\end{figure*}
The convergence of the RCWA method vs number of Fourier harmonics, M,  is illustrated in Fig.\ref{fig2}(d) and (e) on example of $T_{0}$  coefficient. The proposed method provides fast convergence  and results with reasonable accuracy can be already obtained using $M\approx 30-50$ for all considered cases in Fig.\ref{fig2}(d) and (e). In order to quantify the convergence of the RCWA method, we introduce relative error defined as follows $\delta_{M}=|T_{0}^{(M)}-T_{0}^{(200)}|/T_{0}^{(200)}$ (see insets in Fig.\ref{fig2}(d) and (e)). For example, for shallow grating (black circles), $\delta_{30}=0.2\%$ for resonant frequency 1.7 THz and $\delta_{30}=0.01\%$ for the frequency 3.5 THz. Convergence becomes worse for the deep gratings:  at resonant frequency 1.7 THz $\delta_{30}=0.5\%$ (for $h_{g}=1$ \textmu m) and $\delta_{30}=1.3\%$ (for $h_{g}=5$ \textmu m); at non-resonant frequency 3.5 THz $\delta_{30}=0.1\%$ (for $h_{g}=1$ \textmu m) and $\delta_{30}=0.2\%$ (for $h_{g}=5$ \textmu m). Thus, estimations show that convergence of the RCWA method exhibits dependence on grating depth and frequency of the incident radiation.  The cases of the deep gratings and resonant frequencies of the plasmon excitation require account of the larger numbers of Fourier harmonics.

In the case of the plasmonic structure with narrow-slit grating, $w_{g}=0.85$ \textmu m and $a_{g}=1$ \textmu m, our calculations predict much more pronounced features in the optical characteristics including intensity of the plasmon resonances vs grating depth (see Figs.\ref{fig3}). Moreover, narrow-slit grating provides more efficient coupling between incident radiation and plasmon excitations that leads to an emergence of well-pronounced multiple plasmon resonances which are red-shifted in comparison to the previous case. The red-shift of the resonant frequency is the result of a larger contribution of the gated region of 2DEG where phase velocity of the plasmons is smaller than in the ungated region of 2DEG~\cite{Plasmon_res}.

The first plasmon resonance occurs at frequency of $1.38$ THz, at this, absorption of the THz-waves reaches a value of $\sim 50\%$. However, in this spectral range the effect of the grating thickness is still a weak. Starting from the frequencies larger than $2$ THz, spectral characteristics are essentially modified by grating thickness. As seen from Fig.\ref{fig3}(c), deep grating suppresses plasmonic mechanisms of the absorption of THz radiation. The absorption coefficient $L_{0}$ at resonant frequency of $3.78$ THz is decreased from $28\%$ for shallow grating  ($h_{g}=0.05$ \textmu m) to $15\%$
for the deepest grating ($h_{g}=5$ \textmu m). Apparently, this effect relates to an essential increase of the reflectivity of the plasmonic structures with thicker gratings as shown in Fig.\ref{fig3}(b).
It means that for the deeper gratings, a smaller portion of the $em$ energy is concentrated in 2DEG as it will be further illustrated in Section \ref{Sec3}.

It should be noted that application of the RCWA methods for accurate calculations of far-field spectral characteristics of plasmonic structure with narrower-slit grating requires larger
number of Fourier harmonics (see Fig.\ref{fig3}(d) and (e)). Now, the relative errors $\delta_{30}$ for resonant frequency $1.38$ THz are equal to $0.6\%,\, 2.2\%$ and $2.5\%$ for $h_{g}=0.05,\, 1,\,5$ \textmu m, respectively.
The relative errors less than $1\%$ for deep gratings is achievable at $M>60$. Similarly to the previous case, the convergence of the RCWA method is improved at higher frequencies. So, at frequency of $3.78$ THz
accuracy of computation with relative errors $\delta_{M}<1\%$ is achieved at $M>30$.  All spectra shown in Figs.~\ref{fig2} and \ref{fig3} are obtained at $M=100$.

\section{Near-field study}\label{Sec3}

Together with calculations of optical characteristics relating to the far-field, RCWA method allows us to study geometry of the near-field. Especially, we will pay attention to the spatial distributions of the $E_{x}(x,z)$ and
$E_{z}(x,z)$-components of the $em$ fields. Absolute values of these components determine the local absorption of the $em$ wave and can be used for extraction of partial losses in metallic grating and 2DEG.
Having RCWA data on Fourier vectors ${\bf T}_{1}$ (taking from the solutions of master system (\ref{Eq16})) we can find vectors
of the integration constants ${\vec C}^{\pm}_{j}$ in the each j-layer of the structure:
\begin{align}
\left(
\begin{array}{c}
 \vec{C}_{j}^{+}\\
 \vec{C}_{j}^{-}
\end{array}
\right)
=\left(
\begin{array}{c}
{\bf \hat{I}} \\
{\bf {\hat Y}}_{j}{\bf {\hat X}}^{-1}_{j}\hat{\Lambda}_{j}
\end{array}
\right){\bf T}_{j},
\label{Eq28}
\end{align}
where ${\bf T}_{j}$ can be found recurrently ${\bf T}_{j+1}={\bf {\hat X}}^{-1}_{j}\hat{\Lambda}_{j}{\bf T}_{j}$. Substituting found constants into  Eq.~(\ref{Eq10}) (with known ${\bf \hat{W}}_{j}$ and  $\hat{\Lambda}_{j}$ matrices), we can calculate Fourier vectors ${\bf H}_{y,j}(z)$ and  reconstruct a spatial distribution of the $H_{y}$-component in the each $\{x,z\}$ point inside plasmonic structures using Eq. (\ref{Eq3}). The Eqs.  (\ref{Eq14}) and (\ref{Eq15}) are used for reconstruction of the near-field distribution of $H_{y}$ and $E_{x}$ components outside the plasmonic structures.

\begin{figure*}[htbp]
\centering
\includegraphics[width=0.32\textwidth]{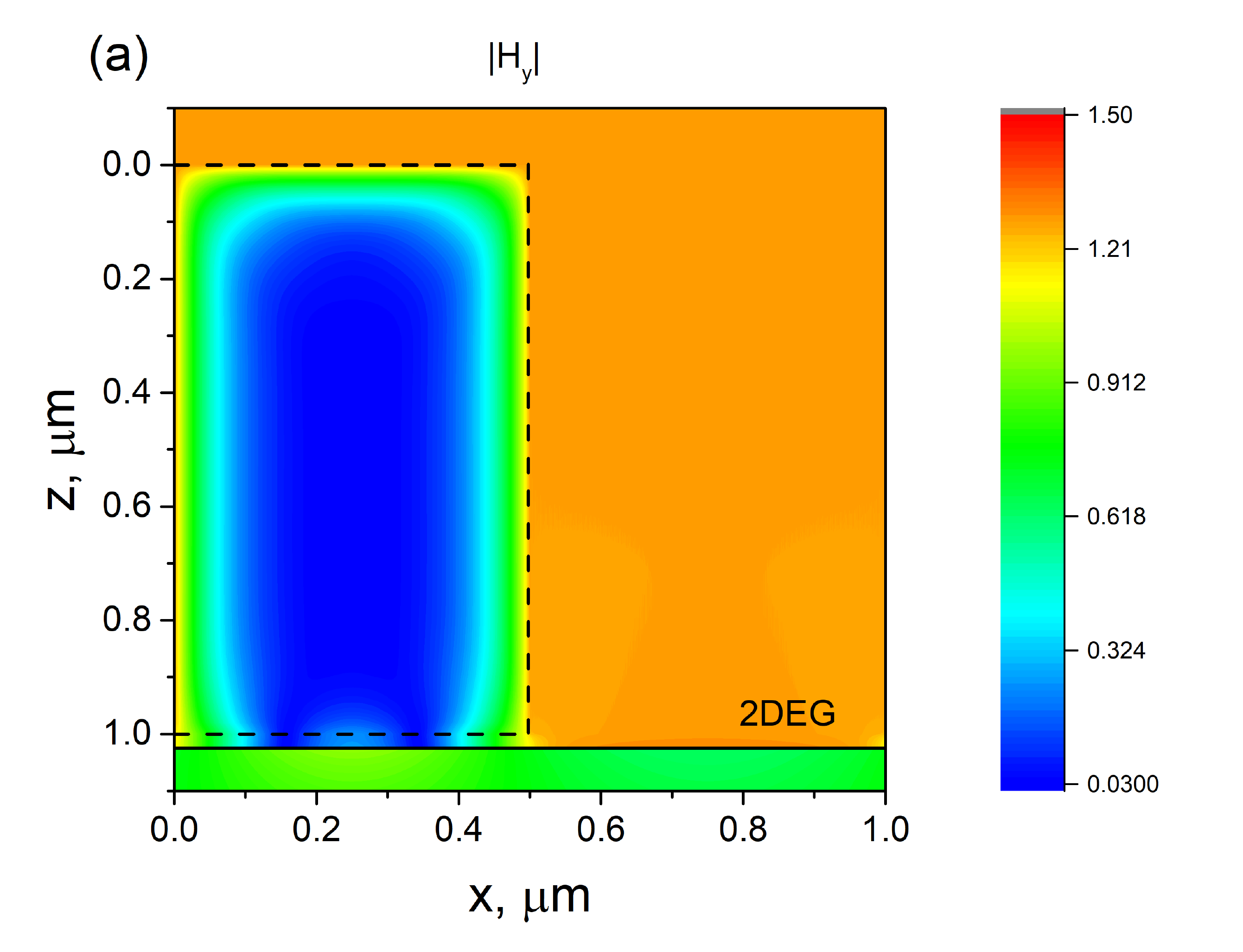}
\includegraphics[width=0.32\textwidth]{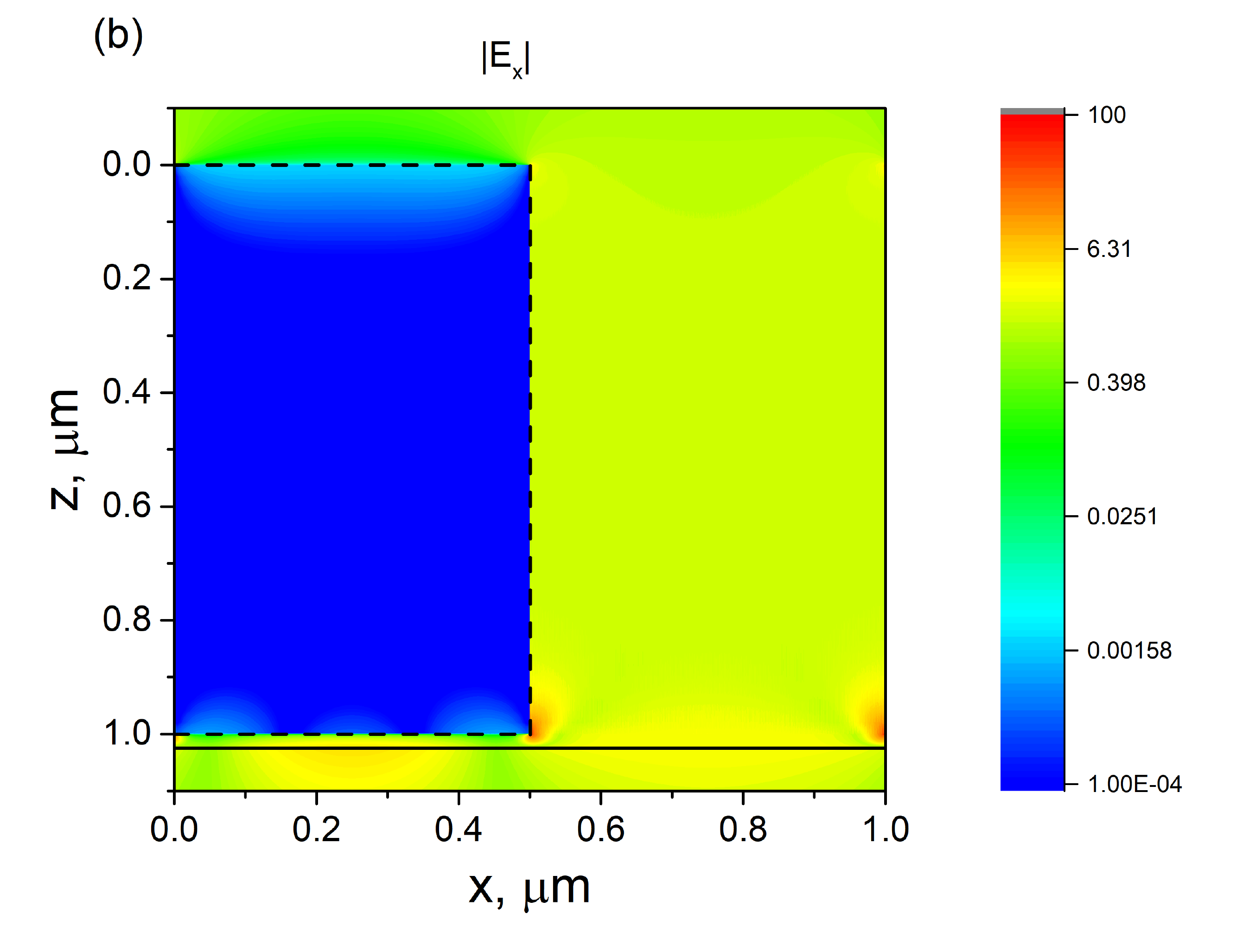}
\includegraphics[width=0.32\textwidth]{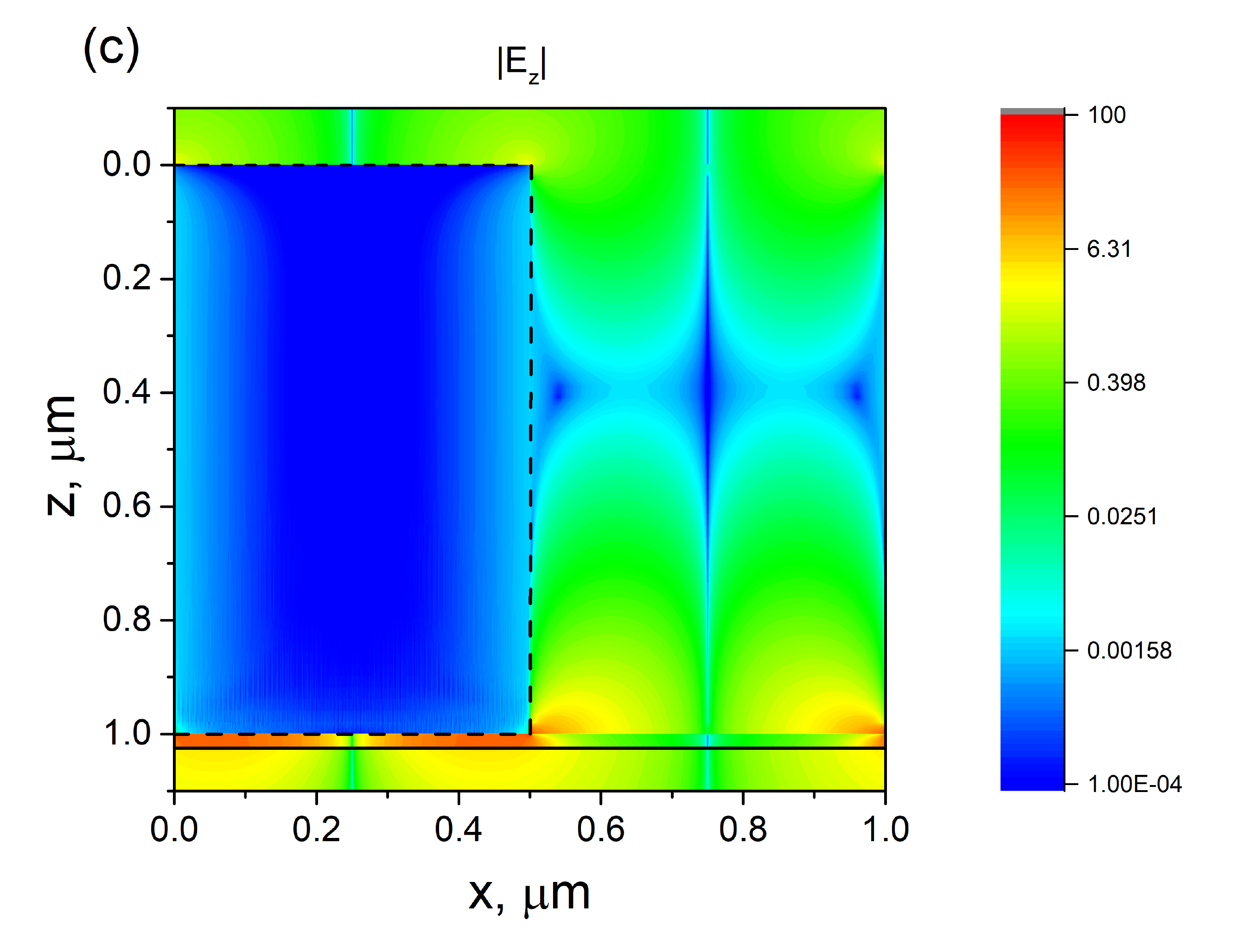}
\caption{The same as in Fig.\ref{fig4} for  $h_{g}=1$ \textmu m.}
 \label{fig5}
\end{figure*}
The spatial distribution of $|H_{y}(x,z)|$ for particular case of the plasmonic structure with shallow grating is shown in Fig.\ref{fig4}(a). This component is tangential to the grating sides and exhibits smooth behavior with partial penetration into the grating bar. Calculations give that skin depth, $\delta=c/\sqrt{2\pi\sigma_{M}\omega}$, of the gold at frequency of $1.7$ THz is equal to  0.057 \textmu m which is comparable with height of the grating bar. The cold zone of $H_{y}$ component occupies middle region of the bar near the bottom face. In the plane of 2DEG, $|H_{y}(x,z)|$ is a discontinuous quantity according boundary conditions (\ref{Eq8}).

The electric components $E_{x}$ (panel(b)) and $E_{z}$ (panel(c)) show more interesting behavior with highly non-uniform distributions. The $E_{z}(x,z)$ can be directly obtained from second relationship in Eqs. (\ref{Eq7}) (using already found Fourier vectors, ${\bf H}_{y,j}(z)$) and Eq. (\ref{Eq3}). For the correct reconstruction of the $E_{x}$ component, we follow the method discussed in Refs.\cite{Brenner, Weismann}.
In the region of the grating, $z\in[0, h_{g}]$,  x-component of the electric field is the normal to the grating bar's sides and one has a discontinuity.
It is more effective to reconstruct a continuous quantity, the component of the displacement field, $D_{x}(x,z)$, which can be easily calculated from the derivative of the $H_{y}$ component with respect to z-coordinate
(see first equation in (\ref{Eq2})). Then $E_{x}(x,y)=D_{x}(x,z)/\epsilon_{1}(x,z)$, where dielectric permittivity $\epsilon_{1}(x,z)$ is the known discontinuous function. Such method allows us partially avoid
an emergence of the unphysical spurious oscillations, known as Gibb's phenomenon. Nevertheless, reconstruction of the near-field patterns requires account of the much more Fourier harmonics than for calculations of the far-field characteristics. This circumstance was discussed in Ref. \cite{Weismann}.

As seen, both $E_{x}$ and $E_{z}$ components demonstrate the field concentration effect. The energy of the $em$ field is mainly concentrated near the ridges (hot zone I) of the metallic bars and in the region between grating bars and 2DEG (hot zone II). In the hot zones (I) and (II) both electric components are essentially enhanced. In the hot zone (II), $E_{z}$ component predominantly dominates. The specific formation of the cold zone for $E_{z}$ component at the vertical axis $x=w_{g}/2$ and $x=(a_{g}+w_{g})/2$  reflects the quadruple-related symmetry of the near field  (for details see Ref. \cite{Korot2014}).
In the hot zones, amplitudes of the electric components can be in several tens times larger than amplitude of incident wave. In spite of the magnetic component, the penetration of the electric components inside metallic bar is strongly suppressed which is result of the edge effects. As seen, $E_x$ component mainly penetrates to the grating's bars from the upper and back faces and $E_{z}$ from the side faces as a tangential ones for corresponding faces.

The similar geometry of the near-fields is realized for the case of the deep grating (see Figs. \ref{fig5}). The mappings of the $E_{x}$ and $E_{z}$ components show that incident wave passes through the subwavelength grating
in the form of TEM mode, i.e in the grating slit, the wave have predominantly polarization along x-direction with almost constant amplitude.

Additionally, we used COMSOL Multiphysics$\textstyle{^\circledR}$\cite{Comsol} to validate independently the obtained results by finite element method. The Wave Optics module \cite{Comsol} is used to solve Maxwell equations for the system, which is shown in Fig.~\ref{fig1}. The 2DEG was introduced as the surface current density at the interface of AlGaN and GaN. The uniform quadratic mesh with 5 nm size is used to resolve near-field components (see Fig.~\ref{fig6add}). The COMSOL's results of the electric components distributions for the case of the shallow grating with $h_g=0.05$ \textmu m are shown in Fig.~\ref{fig6add}.  Excellent agreement between the modified RCWA and finite element methods is demostrated in both far- and near field studies.

\begin{figure}[htbp]
\centering
\includegraphics[width=0.32\textwidth]{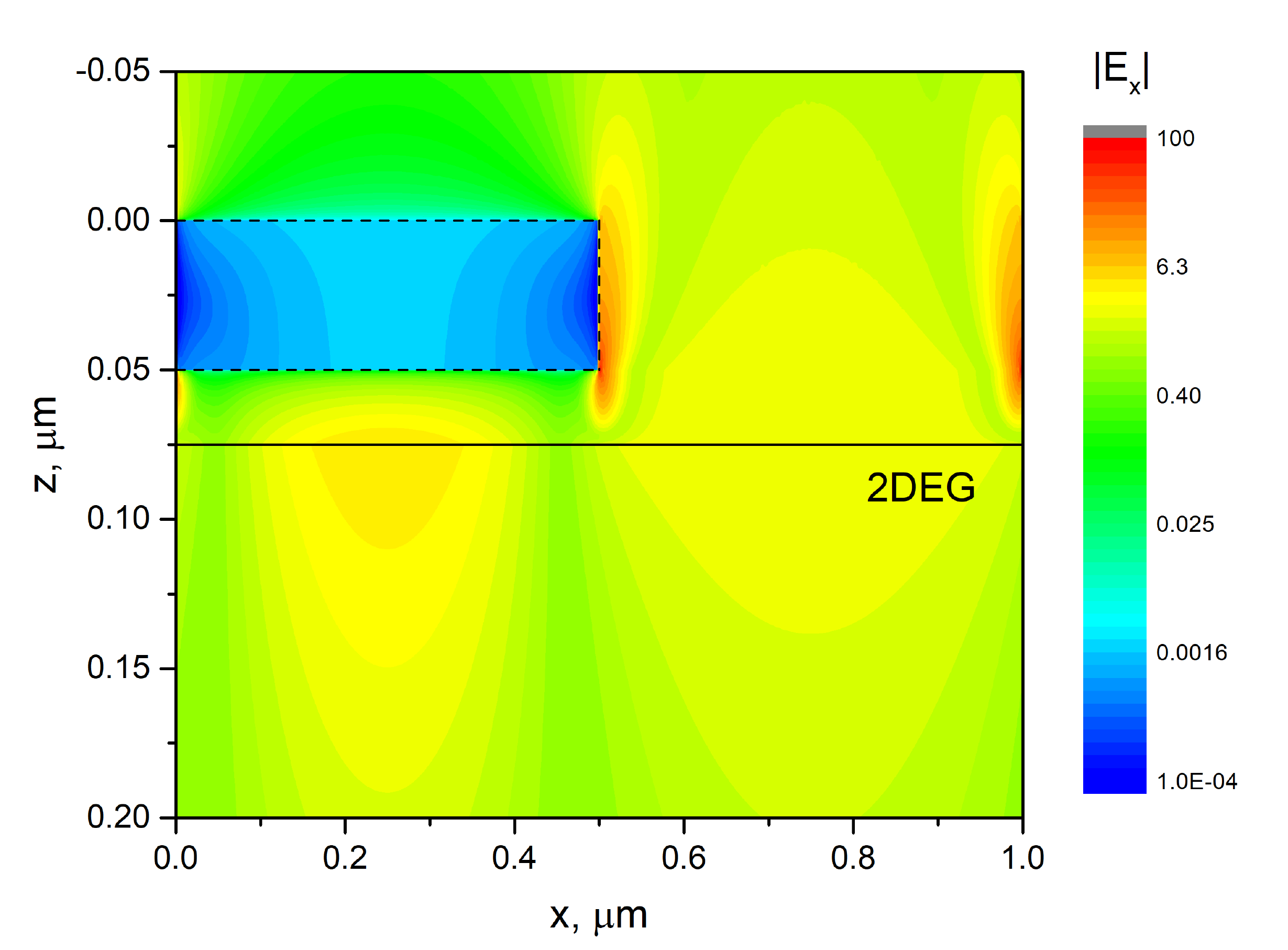}
\includegraphics[width=0.32\textwidth]{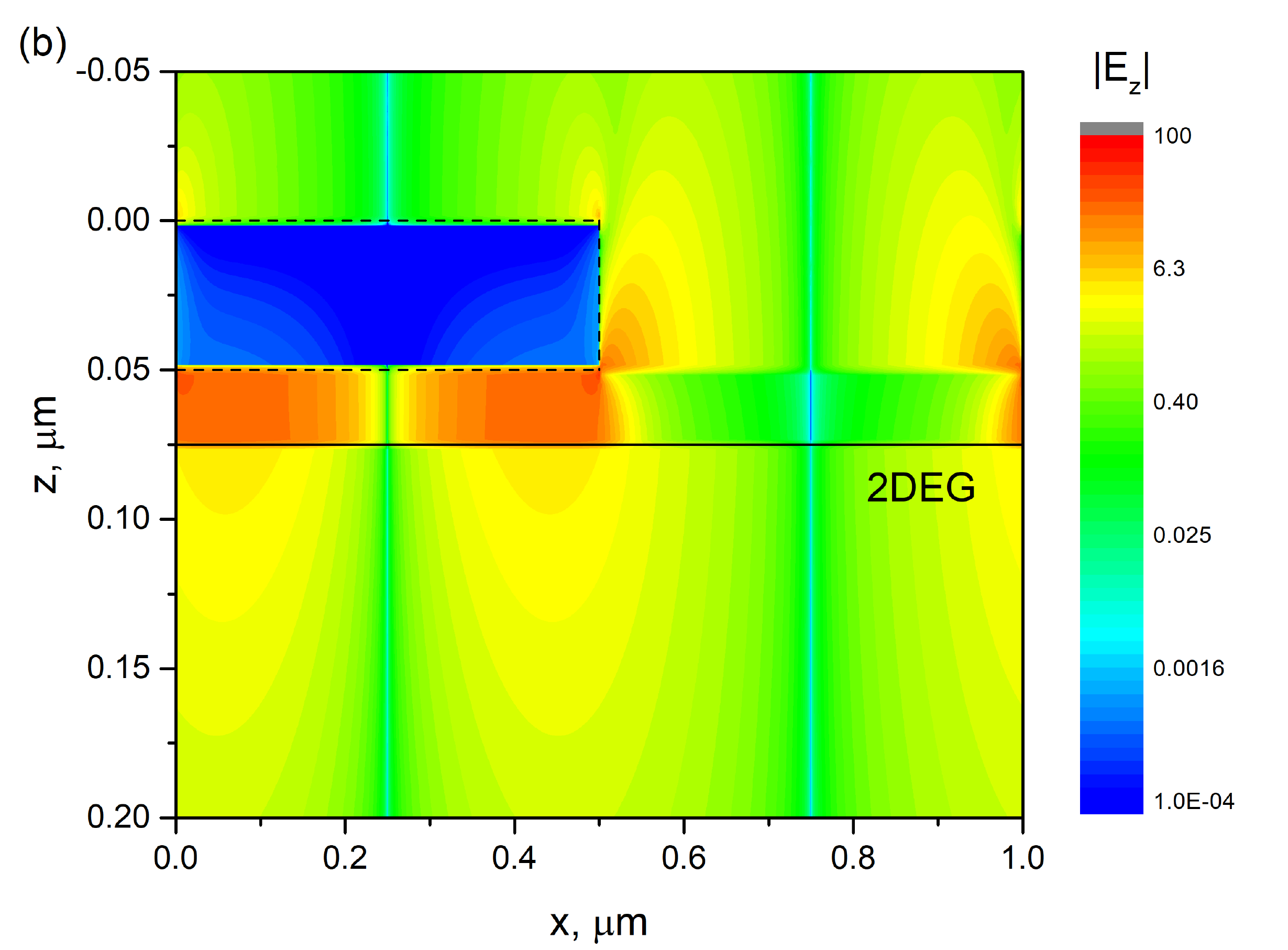}
\caption{The same as in Fig.\ref{fig4} obtained by usage of COMSOL Multiphysics. }
 \label{fig6add}
\end{figure}

The spatial distributions of the $|E_{x}|(x,z)$ and $|E_{z}|(x,z)$ can be used for calculation of the partial losses in the grating, $L_{gr}$
and 2DEG, $L_{2DEG}$:
\begin{equation}
L_{gr}=\frac{4\pi\sigma_{M}}{c}\frac{\int_{0}^{w_{g}}\int_{0}^{h_{g}}dxdz|E_{x}|^2+|E_{z}|^2}{\int_{0}^{a_{g}}dx|E_{ins}|^2}
\end{equation}
and
\begin{equation}
L_{2DEG}=\frac{4\pi\text{Re}[\sigma^{2D}_{\omega}]}{c}\frac{\int_{0}^{a_{g}}dx|E_{x}(x,z_{2})|^2}{\int_{0}^{a_{g}}dx|E_{ins}|^2}.
\label{Abs}
\end{equation}
where $E_{ins}$ is the amplitude of the incident wave.

For the case in Fig.\ref{fig4}, we obtained that $L_{gr}=0.184\%$
that consists of $0.0741\%$ contribution of $E_{x}$-component and $0.11\%$ contribution of $E_{z}$-component. Losses in 2DEG are considerably larger, $L_{2DEG}=37.92\%$. Total losses from the near-field patterns, $L_{0}=L_{2DEG}+L_{gr}=38.11\%$. Calculations of the $L_{0}$ from far-field characteristics gives the almost same value $38.21\%$. For the case of deep grating (see Fig.\ref{fig5}), we obtained the increase of absorption in the grating bars, $L_{gr}=0.56\%$, (with $0.072\%$ and $0.49\%$ contributions for $E_{x}$ and $E_{z}$ components, respectively) with almost same value of absorption in 2DEG $L_{2DEG}=37.97\%$. The total losses  $L_{0}=38.53\%$ that almost coincide with the number, $38.58\%$  obtained from  far-field characteristics.

Calculations of the partial losses at the frequency of the 1-st order plasmon resonance indicate that incident $em$ wave is mainly absorbed by 2DEG and this  absorption weakly depends on thickness of grating bars. This fact is illustrated  by the spatial distribution of the amplitude of the $x-$component of the electric field, $|E_{x}(x,z_{2})|$,  in the plane of 2DEG, calculated at the frequencies of the 1-st (Fig.~\ref{fig6}(a)) and 2-nd (Fig.~\ref{fig6}(b)) plasmon resonances at three values of the grating depth. As seen, all three (grey, red, green) curves for lower frequency ($\omega/2\pi=1.7$ THz) almost coincide and all of them exhibit non uniform, oscillating-like behavior in the gated region with almost flat distribution in the ungated region. The obtained distribution denotes that larger part of $em$ energy is absorbed in the gated region i.e. under metallic strip.

\begin{figure}[t!]
\centering
\includegraphics[width=0.4\textwidth]{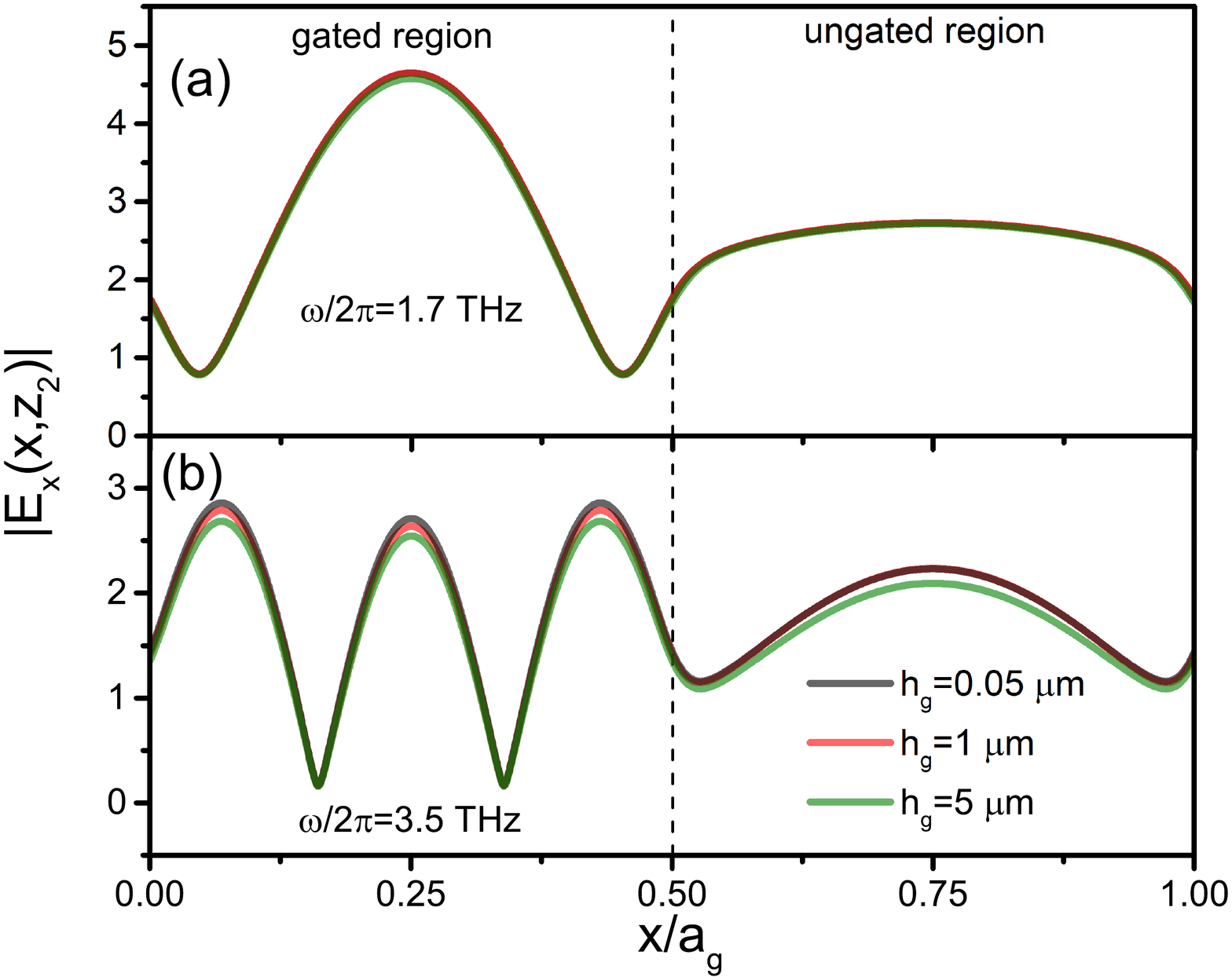}
\caption{(color online): Distribution of $|E_{x}(x,z_{2})|$ on one spatial period of the plasmonic structure with $w_{g}=0.5$ \textmu m and $a_{g}=1$ \textmu m at two resonant frequencies.  }
 \label{fig6}
\end{figure}

For the higher frequency ($\omega/2\pi=3.5$ THz), spatial distribution of the $|E_{x}(x,z_{2})|$ quantity (Fig.\ref{fig6}(b)) acquires more complicated form with several spatial oscillations in the gated region. At this, the effect of the grating thickness becomes visible, i.e, the deep gratings starts to screen the interaction of $em$ wave with 2DEG.
The emergence of the several spatial oscillations in  distribution of $|E_{x}(x,z_{2})|$ leads to suppression of absorptivity of the plasmonic structures at higher order plasmon resonances. Also, according Drude model, at higher frequencies response of electron gas on $em$ wave becomes weaker which leads to a decrease of the prefactor standing in Eq.\ref{Abs}. This prefactor, $4\pi\text{Re}[\sigma^{2D}_{\omega}]/c$, for two considered frequencies $\omega/2\pi=1.7$ and $\omega/2\pi=3.5$ THz is equal to $0.049$ and $0.012$, respectively.   Using the obtained distributions in Fig.\ref{fig6}(b), we found that for $h_{g}=0.05,\,1,\,5$ \textmu m, $L_{2DEG}=4.42,\,4.3$ and $3.8\,\%$ and the corresponding values of total losses calculated from far-field characteristics, $L_{0}=4.8,\,5.3,\,6.3\%$. Note, that for the structure with deepest grating the absorptions in 2DEG and grating bars become comparable.

The plasmonic structure with narrow-slit grating provides more efficient coupling between 2DEG and $em$ radiation. The distributions of $|E_{x}(x,z_{2})|$ calculated for the structure with $w_{g}=0.85$ \textmu m at two resonant frequencies $\omega/2\pi=1.38$ THz (1-st order plasmon resonance) and $\omega/2\pi=3.78$ THz (3-rd order plasmon resonance) are shown in Figs. \ref{fig7}. As seen, the geometry of the distributions obtained for the frequency of
1-st order plasmon resonance (Fig.\ref{fig7}(a)) is similar to the previous case depicted in Fig.\ref{fig6}(a). However, the wider gated region of 2DEG integrally provides larger contribution to the absorption of $em$ wave by 2DEG. The corresponding values of $L_{2DEG}$ are following: $47.5\%$ (for $h_g=0.05$ \textmu m), $46.6\%$ (for $h_g=1$ \textmu m) and $41.2\%$ (for $h_g=5$ \textmu m). At this, $L_{0}=48.2,\,47.6,\,43.9\%$, respectively.
The distributions in Fig.\ref{fig7}(b) obtained at the frequency of 3-rd order plasmon resonance demonstrate multiple spatial oscillations. The  number of such oscillations is proportional to the order of plasmon resonances. Also, we see that deepest grating with $h_g$=5 \textmu m
essentially suppresses the plasmon absorption of $em$ wave. The corresponding values of $L_{2DEG}$ are following: $27.2\%$ (for $h_g=0.05$ \textmu m), $24.1\%$ (for $h_g=1$ \textmu m) and $11.8\%$ (for $h_g=5$ \textmu m). At this, $L_{0}=28.7,\,26.1,\,15.1\%$, respectively.

\begin{figure}[t!]
\centering
\includegraphics[width=0.4\textwidth]{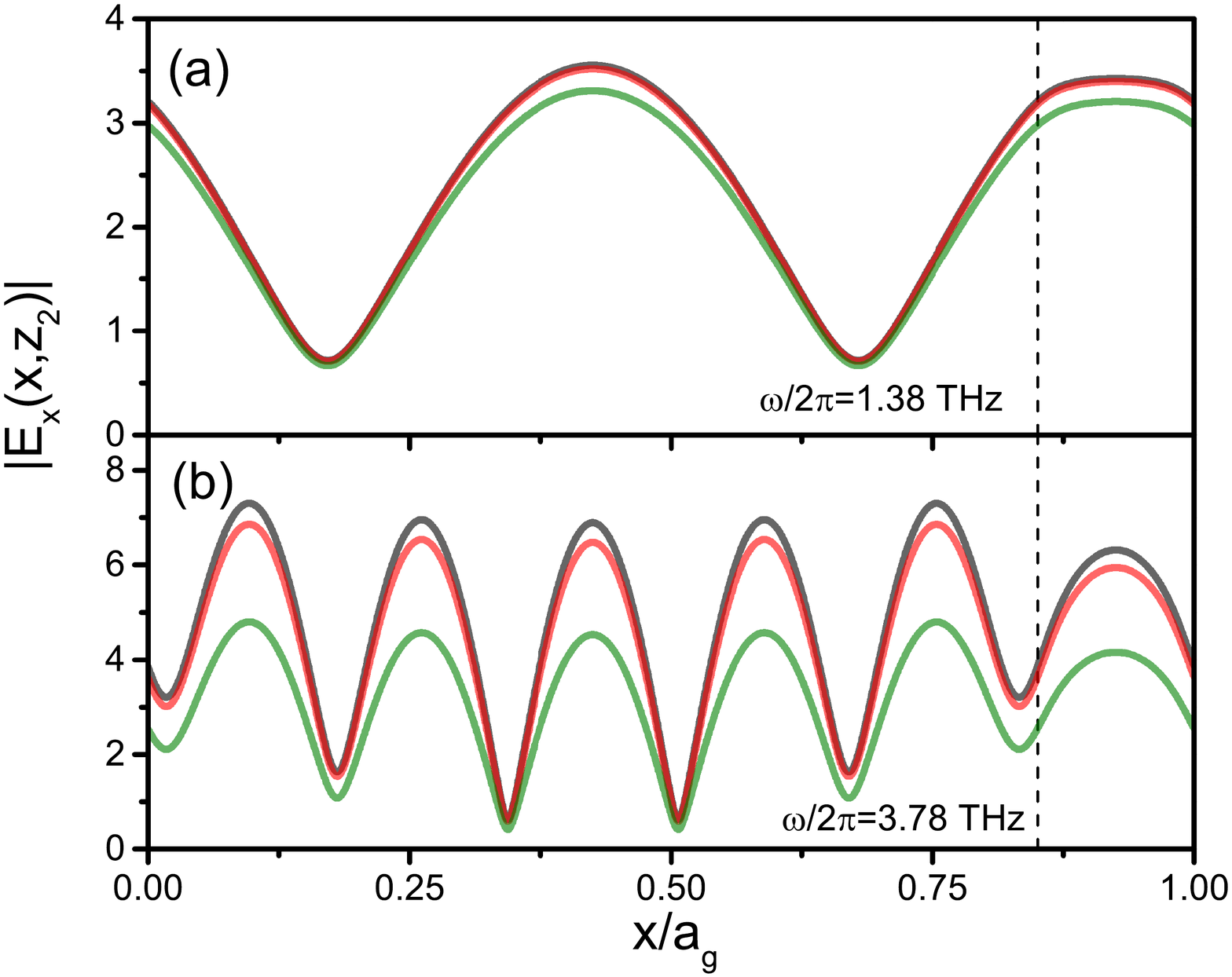}
\caption{(color online): The same as in Fig.\ref{fig6} for  $w_{g}=0.85$ \textmu m.}
 \label{fig7}
\end{figure}

\section{Summary} \label{Sec4}
We have developed computationally stable RCWA method for solution of Maxwell's equations in the case of the multi-layered plasmonic structures with delta-thin grating-gated conductive channel. The method was formulated for planar diffraction problem for both TM and TE polarization of incident wave. Method was implemented for investigation of far- and near-field characteristics of the particular plasmonic structures based on AlGaN/GaN heterostructure with deeply subwavelength metallic grating coupler.

The calculations of the far-field characteristics including transmission, reflection and absorption coefficients for zero diffraction order were performed in THz frequency range where considered structure has multiple resonances related to the excitations of 2D plasmons in conductive channel of AlGaN/GaN heterostructure. The dependence of these characteristics vs grating parameters and their convergence vs number of the Fourier harmonics were analyzed.

We found that spectra of transmission and reflection coefficients in the lower frequency range, $0..2$ THz, have weak dependence on grating depth. Results for both shallow ($h_{g}/a_{g}=0.05$ \,\textmu m/$1$ \textmu m) and deep ($h_{g}/a_{g}=1$,\, $5$\, \textmu \text{m}/$1$ \textmu m)  gold grating are almost identical and coincide with the results of IE method where grating is treated as delta-thin. In higher frequency range, $3..5$ THz, increase of the grating depth suppress transmission with increasing of the reflection coefficients. At the same time, absorption spectrum remain less sensitive to the grating depth. We showed that dispersion of far field characteristics on $h_{g}$ becomes more pronounced for narrower-slit grating with $w_{g}/a_{g}=0.85$ \,\textmu m/$1$ \textmu m
than for wide-slit grating with $w_{g}/a_{g}=0.5$ \,\textmu m/$1$ \textmu m. We showed that convergence of the calculations depends on geometrical parameters of the grating and frequencies. Better convergence is achieved for shallow and wide-slit grating with relative errors of $\sim 0.05-0.2\%$ (in dependence on frequency) with  $30$ Fourier harmonics.  For deep and narrow-slit grating, relative errors of $\sim 0.1..1 \%$ is achieved at $M\sim 50$.

Procedure of the calculations of the near-field characteristics was discussed in detail. Analysis of spatial distribution of the amplitudes of the $em$ wave's components in the near field-zone reveals the main physical peculiarities of the interaction of the plasmonic structure with incident radiation. It was shown that  subwavelength metallic grating plays a role of perfect waveguide for incident wave,  concentrator of the $em$ energy and polarization rotator.
In the region of the grating slit $em$ wave has predominantly lateral polarization with amplitude close to amplitude of incident wave.
The hot zone is formed in region between grating bars and 2DEG where $em$ wave has predominantly vertical polarization with amplitudes that can in $100$ times exceed the amplitude of incident wave.

The pattern of the near-field also was used for the calculations of the partial losses related to the grating and 2DEG.
It was shown that at the frequencies of the plasmon resonances the structure can efficiently absorb THz radiation with absorption coefficient values in the order of $20-50\,\%$ (in dependence of the grating filling factor and order of plasmon resonance). We found that contribution of 2DEG to the total losses is dominant at low-frequency plasmon resonances with weak dependence on the grating depth. At high-frequency plasmon resonances, the effect of the grating depth becomes essential. The deep gratings can effectively screen interaction of the $em$ waves with plasmon oscillation in 2DEG that leads to an decrease of the total absorption of THz radiation.

Also, it should be noted that the proposed modified RCWA has several advantages over conventional {\it volumetric} RCWA. First one, our realization of RCWA method allows us to avoid additional numerical manipulation with matrices that can reduce the computational time.   For considered structure, we have 25$\%$ in term of computation time savings in comparison with conventional RCWA at the volumetric treatment of conductive layer.
This value can be increased in simulation of structures with the stack of 2D conductive layers. Second one, we operate with one parameter, two-dimensional concentration, $n_{2D}$, instead of two independent parameters of bulk concentration, $n_{3D}$, and thickness of the layer, $d$. It can be convenient for metrology of the structures at the processing of the experimental data.

We suggest that proposed modified RCWA algorithm can be effectively used for the modeling of the optical characteristics of various kinds of plasmonic structures
with 2D conductive channels, including quantum wells- or graphene-based structures and results of the paper provide deeper insight on physics of the interaction of THz radiation with grating-gated plasmonic structures.

\section{Funding}
The work was supported by the Research Council of Lithuania (Lietuvos mokslo taryba)
under the "KOTERA-PLAZA" Project (Grant No. DOTSUT-247) funded by the European Regional Development Fund according to the supported activity "Research Projects Implemented by World-class Researcher Groups" under Measure No. 01.2.2-LMT-K-718-0047. SMK was supported by the Bundesministerium f\"{u}r Bildung und Forschung through the project VIP+ "Nanomagnetron".

\section{Acknowledgments}
Authors thanks to Prof. V. A. Kochelap (ISP NASU, Ukraine) and Dr. I. Ka\v{s}alynas (FTMC, Lithuania) for fruitful discussions of the various aspects of this work.

\medskip

\noindent\textbf{Disclosures.} The authors declare no conflicts of interest.




\begin{thebibliography}{1}
\bibitem{Popov_Grating} E. Popov ed. \enquote{Gratings: Theory and Numeric Applications}, First Edition, Presses universitaires de Provence (PUP), (2012).
\bibitem{Neumann} W. Neumann, \enquote{Fundamentals of Dispersive Optical Spectroscopy Systems}, SPIE Press, Bellingham, Washington, USA (2014).
\bibitem{Antenna} H. F. Hammad, Y. M. M. Antar, A. P. Freundorfer and M. Sayer, \enquote{A new dielectric grating antenna at millimeter wave frequency,} IEEE Transactions on Antennas and Propagation, \textbf{52}, 36-44, (2004).
\bibitem{Polarizers} S. Shena, Y. Yuana, Z. Ruana, and H. Tan, \enquote{Optimizing the design of an embedded grating polarizer for infrared polarization light field imaging,} Results in Physics, \textbf{12}, 21-31 (2019).
\bibitem{Shaligin2016} G. A. Melentev, V. A. Shalygin, L. E. Vorobjev, V. Yu. Panevin, D. A. Firsov, L. Riuttanen, S. Suihkonen, V. V.
Korotyeyev, Yu. M. Lyaschuk, V. A. Kochelap, and V. N. Poroshin, \enquote{Interaction of surface plasmon polaritons in heavily doped GaN microstructures with
terahertz radiation,} J. Appl. Phys. \textbf{119}, 093104 (2016).
\bibitem{Vitovt2020} V. Janonis, S. Tumenas, P. Prystawko, J. Kacperski, and I. Ka\v{s}alynas, \enquote{Investigation of n-type gallium nitride grating
for applications in coherent thermal sources,} Appl. Phys. Lett. \textbf{116}, 112103 (2020).
\bibitem{Popov} V. V. Popov, D. V. Fateev, O. V. Polischuk, and M. S. Shur, \enquote{Enhanced electromagnetic coupling between terahertz radiation and plasmons in a grating-gate transistor structure on membrane substrate,} Opt. Express \textbf{18}, 16771 (2010).
\bibitem{Korot2018} V. V. Korotyeyev, V. A. Kochelap, S. Danylyuk, and L. Varani,  \enquote{Spatial dispersion of the high-frequency conductivity of two-dimensional electron gas subjected to a high electric field: Collisionless case,} Appl. Phys. Lett. \textbf{113}, 041102  (2018).
\bibitem{Rizhii2020} V. Ryzhii, T. Otsuji, and M. Shur, \enquote{Graphene based plasma-wave devices for terahertz applications}, Appl. Phys. Lett. \textbf{116}, 140501 (2020).
\bibitem{Otsuji2014} T. Otsuji and M. Shur, \enquote{Terahertz Plasmonics: Good Results and Great Expectations,} IEEE Microw. Mag. \textbf{15}, 43-50 (2014).
\bibitem{Pashnev2020} D. Pashnev, T. Kaplas, V. Korotyeyev, V. Janonis, A. Urbanowicz, J. Jorudas and I. Ka\v{s}alynas, \enquote{Terahertz time-domain spectroscopy of
two-dimensional plasmons in AlGaN/GaN heterostructures,} Appl. Phys. Lett. \textbf{117}, 051105 (2020).
\bibitem{Pashnev2020b} D. Pashnev, V. Korotyeyev, V. Janonis, J. Jorudas, T. Kaplas, A. Urbanowicz, and I. Ka\v{s}alynas, \enquote{Experimental evidence of temperature
dependent effective mass in AlGaN/GaN heterostructures observed via THz spectroscopy of 2D plasmons} Appl. Phys. Lett. \textbf{117}, 162101 (2020).
\bibitem{Yan2015} Bo Yan, Jingyue Fang, Shiqiao Qin, Yongtao Liu, Yingqiu Zhou, Renbing Li and Xue-Ao Zhang \enquote{Experimental study of plasmon in a grating coupled graphene device
with a resonant cavity} Appl. Phys. Lett. \textbf{107}, 191905 (2015)
\bibitem{Jadidi2015} M. M. Jadidi, A. B. Sushkov, R. L. Myers-Ward, A. K. Boyd, K. M. Daniels, D.K. Gaskill,  M. S. Fuhrer, H. Dennis Drew and T. E. Murphy \enquote{Tunable Terahertz Hybrid Metal - Graphene Plasmons}
Nano Lett. {\textbf 15}, 7099 -7104 (2015).
\bibitem{Zhao2015} Bo Zhao and Zhuomin M. Zhang \enquote{Strong Plasmonic Coupling between Graphene Ribbon Array and Metal Gratings} ACS Photonics \textbf{2}, 1611 -1618 (2015).
\bibitem{Lu2016} Hua Lu, Jianlin Zhao and Min Gu \enquote{Nanowires-assisted excitation and propagation of mid-infrared surface plasmon polaritons in graphene} J. Appl. Phys. {\bf 120}, 163106 (2016).
\bibitem{Kukhtaruk} S.M. Kukhtaruk, V.V. Korotyeyev, V.A. Kochelap and L. Varani, \enquote{Interaction of THz Radiation with Plasmonic Grating
Structures Based on Graphene,} Proceedings of International Conference on Mathematical Methods in Electromagnetic Theory, Lviv, pp. 196-199 (2016).
\bibitem{Low} T. Low, A. Chaves, J. D. Caldwell, A. Kumar, N. X. Fang, Ph. Avouris, T. F. Heinz, F. Guinea, L. Martin-Moreno,  F. Koppens, \enquote{Polaritons in layered two-dimensional materials,} Nature Materials \textbf{16}, 182-194 (2017).
\bibitem{Michailov} S.A. Mikhailov, \enquote{Plasma instability and amplification of electromagnetic waves in low-dimensional electron systems,}
Phys. Rev. B \textbf{58}, 1517 (1998).
\bibitem{Korot2020} V. V. Korotyeyev and V. A. Kochelap, \enquote{Plasma wave oscillations in a nonequilibrium two-dimensional electron gas:
Electric field induced plasmon instability in the terahertz frequency range,} Phys. Rev. B \textbf{101}, 235420 (2020).
\bibitem{Popov2010} D. V. Fateev, V. V. Popov, and M. S. Shur, \enquote{Plasmon spectra transformation in grating-gate transistor structure with spatially modulated two-dimensional electron channel},  Semiconductors \textbf{44}, 1455 (2010) [Fiz. Tekh. Poluprovodn. (St. Petersburg) \textbf{44}, 1455-1462 (2010)].
\bibitem{Popov2011} V. V. Popov, D. V. Fateev, T. Otsuji, Y. M. Meziani, D. Coquillat, and W. Knap, \enquote{Plasmonic terahertz detection by a double-grating-gate field-effect transistor structure with an asymmetric unit cell,} Appl. Phys. Lett. \textbf{99}, 243504 (2011).
\bibitem{Korot2017} Y. M. Lyaschuk and V. V. Korotyeyev, \enquote{Theory of detection of terahertz radiation in hybrid plasmonic structures with drifting electron gas,} Ukr. J. Phys. \textbf{62}(10), 889 (2017).
\bibitem{Nosich2013} Olga V. Shapoval, Juan Sebastian Gomez-Diaz, Julien Perruisseau-Carrier, Juan R. Mosig  and Alexander I. Nosich, IEEE Transactions on Terahertz Science and Technology
\enquote{Integral Equation Analysis of Plane Wave Scattering by Coplanar Graphene - Strip Gratings in the THz Range} \textbf{3}(5), 666-674 (2013).
\bibitem{Nosich2013Aip} O. V. Shapovala and A. I. Nosich \enquote{Finite gratings of many thin silver nano strips: Optical resonances and role of periodicity,} AIP Advances {\bf 3}, 042120 (2013).
\bibitem{Moharam} M. G. Moharam and T. K. Gaylord, \enquote{Rigorous coupled-wave analysis of planar-grating diffraction,} J. Opt. Soc. Am \textbf{71}(7), 811-818 (1981).
\bibitem{Moharam1995} M. G. Moharam, E. B. Grann, D. A. Pommet and T. K. Gaylord, \enquote{Formulation for stable and efficient implementation of the rigorous coupled-wave analysis of binary gratings,}  J. Opt. Soc. Am. A  \textbf{ 12}(5) 1068-1076 (1995).
\bibitem{Gaylord} M. G. Moharam, Drew A. Pommet, Eric B. Grann and T. K. Gaylord, \enquote{Stable implementation of the rigorous coupled-wave
analysis for surface-relief gratings: enhanced transmittance matrix approach,} J. Opt. Soc. Am. A \textbf{12}(5), 1077-1086 (1995).
\bibitem{Inampudi2017} S. Inampudi and H. Mosallaei \enquote{Tunable wideband-directive thermal emission from SiC surface using bundled graphene sheets}, Phys. Rev B {\bf 96}, 125407 (2017).
\bibitem{Inampudi2019} S. Inampudi, V. Toutam and S. Tadigadapa \enquote{Robust visibility of graphene monolayer on patterned plasmonic substrates}, Nanotechnology {\bf 30}, 015202 (2019).
\bibitem{Li} L. Li, \enquote{Use of Fourier series in the analysis of discontinuous periodic structures,} J. Opt. Soc. Am. A \textbf{13}(9), 1870- 1876 (1996).
\bibitem{Popov2004} E. Popov, M. Nevi\`{e}re, and N. Bonod, \enquote{Factorization of products of discontinuous functions applied to Fourier-Bessel basis,} J. Opt. Soc. Am. A \textbf{21}(1), 46-52 (2004).
\bibitem{Blesz1993} E. Bleszynski, M. Bleszynski and T. Jaroszewicz \enquote{Surface-Integral Equations for Electromagnetic Scattering from Impenetrable and Penetrable Sheets,}
IEEE Antennas and Propagation Magazine \textbf{35}(6), 14-25 (1993).
\bibitem{Nosich1998} T. L. Zinenko and A. I. Nosich \enquote{Plane Wave Scattering and Absorption by Resistive-Strip and Dielectric-Strip Periodic Gratings,}
IEEE Transactions on Antennas and Propagation \textbf{46}(10), 1498-1505 (1998).
\bibitem{Gusynin2007} V. P. Gusynin, S. G. Sharapov, and J. P. Carbotte, \enquote{Sum rules for the optical and Hall conductivity in graphene,} Phys. Rev. B {\textbf 75}, 165407 (2007).
\bibitem{Falkovsky}   L. A. Falkovsky and S. S. Pershoguba, \enquote{Optical far-infrared properties of a graphene monolayer and multilayer,} Phys. Rev B \textbf{76}, 153410 (2007).
\bibitem{Balaban2013} M. V. Balaban, O. V. Shapoval, and A. I. Nosich \enquote{THz wave scattering by a graphene strip and a disk in the free space: integral equation analysis and surface plasmon resonances,}
 J. Opt. {\textbf 15}, 114007 (2013).
  \bibitem{Kukhtaruk2016} S.M. Kukhtaruk, V.A. Kochelap, V.N. Sokolov, K.W. Kim, \enquote{Spatially dispersive dynamical response of hot carriers in doped graphene,} Physica E: Low-dimensional Systems and Nanostructures, \textbf{ 79}, 26-37, (2016).
\bibitem{Korot2014}  Y. M. Lyaschuk and V. V. Korotyeyev, \enquote{Interaction of a Terahertz electromagnetic wave with the plasmonic system "grating-2D-gas". Analysis of features of the near field,}  Ukr. J. Phys. \textbf{59}(5), 495-504 (2014).
\bibitem{Chaplik} A. V. Chaplik, Surf. Sci. Rep. \enquote{Absorption and emission of  electromagnetic waves by two-dimensional plasmons} {\bf 5}, 289 (1985).
\bibitem{DyakonovPRL93} M. Dyakonov and M. Shur, \enquote{Shallow water analogy for a ballistic fild effct transistor: New mechanism of plasma wave
generation by dc current} Phys. Rev. Lett., {\bf 71}, 2465-2467 (1993).
\bibitem{Plasmon_res} Parametrical studies of the plasmon resonance at different configurations of grating coupler is reported in Ref.\cite{Korot2014} and recent experimental paper Ref.\cite{Pashnev2020}.
\bibitem{Brenner} K.-H. Brenner, \enquote{Aspects for calculating local absorption with the rigorous coupled-wave method,} Optics Express \textbf{18}(10),  10369 (2010).
\bibitem{Weismann} M. Weismann, Dominic F G Gallagher, and Nicolae C Panoiu, \enquote{Accurate near-field calculation in the rigorous coupled-wave analysis method,} Journal of Optics, \textbf{17}(12), 125612 (2015).
\bibitem{Comsol} COMSOL Multiphysics$\textsuperscript{\textregistered}$~v.~5.5.~www.comsol.com. COMSOL AB, Stockholm, Sweden.
\end{thebibliography}
\end{document}